\documentclass[aps, prd, twocolumn, superscriptaddress, nofootinbib]{revtex4}
\usepackage{graphicx}
\usepackage{dcolumn}
\usepackage{amssymb}
\usepackage{amsmath}%
\usepackage{color}
\usepackage[normalem]{ulem}
\usepackage{float}

\graphicspath{{figures/}}

 \newenvironment{itemz}
 {\begin{list}{$\bullet$}{\setlength{\itemsep}{0pt}}}
 {\end{list}}

\newcommand{\bitz}{\begin{itemz}}
\newcommand{\eitz}{\end{itemz}}

\def\al{\alpha}  
\def\be{\beta} 
\def\ga{\gamma}

\def\th{\theta}

\def\ka{\kappa}
\def\la{\lambda}

\def\si{\sigma}
\def\ta{\tau}

\def\bk{{\mathbf{k}}}

\def\bD{{\mathbf{D}}}

\def\mcL{{\mathcal L}}

\newcommand{\ben}{\begin{equation}}
\newcommand{\een}{\end{equation}}
\newcommand{\bea}{\begin{eqnarray}}
\newcommand{\eea}{\end{eqnarray}}
\newcommand{\ba}{\begin{array}}
\newcommand{\ea}{\end{array}}
\newcommand{\bit}{\begin{itemize}}
\newcommand{\eit}{\end{itemize}}

\def\math{\mathsurround 0pt}
\def\oversim#1#2{\lower.5pt\vbox{\baselineskip0pt \lineskip-.5pt
        \ialign{$\math#1\hfil##\hfil$\crcr#2\crcr{\scriptstyle\sim}\crcr}}}

\def\pa{\partial}

\def\half{\frac{1}{2}}

\newcommand{\vev}[1]{\left\langle#1\right\rangle}

\newcommand{\corrLen}{\ell_\text{c}}

\newcommand{\Slen}{\ell} 

\newcommand{\SwidS}{r_\text{s}} 
\newcommand{\SwidG}{r_\text{g}} 

\newcommand{\tStart}{\tau_\text{start}}
\newcommand{\tDiff}{\tau_\text{diff}}

\newcommand{\tCG}{\tau_\mathrm{cg}}
\newcommand{\tRef}{\tau_\mathrm{ref}}

\newcommand{\tEnd}{\tau_\text{end}}

\newcommand{\tSim}{\tau_\text{sim}}

\newcommand{\xiWind}{\xi_\text{w}}
\newcommand{\xiLag}{\xi_\mcL}

\newcommand{\ie}{\textit{i.e.}}

\newcommand{\rtil}{\tilde{r}}

\begin{document}


\title{Type I Abelian Higgs strings: evolution and Cosmic Microwave Background constraints}

\newcommand{\addressSussex}{Department of Physics \& Astronomy, University of Sussex, Brighton, BN1 9QH, United Kingdom}
\newcommand{\HIPetc}{\affiliation{
		Department of Physics and Helsinki Institute of Physics,
		PL 64, 
		FI-00014 University of Helsinki,
		Finland
	}}

\author{Mark Hindmarsh} 
\email{m.b.hindmarsh@sussex.ac.uk}
\affiliation{\addressSussex}
\HIPetc

\author{Joanes Lizarraga}
\email{joanes.lizarraga@ehu.eus}
\affiliation{Department of Theoretical Physics, University of the Basque Country UPV/EHU, 48080 Bilbao, Spain}

\author{Jon Urrestilla}
\email{jon.urrestilla@ehu.eus}
\affiliation{Department of Theoretical Physics, University of the Basque Country UPV/EHU, 48080 Bilbao, Spain}

\author{David Daverio}
\email{dd415@cam.ac.uk}
\affiliation{Centre for Theoretical Cosmology, Department of Applied Mathematics and Theoretical Physics, Wilberforce Road, Cambridge CB3 0WA, United Kingdom}
\author{Martin Kunz}
\email{martin.kunz@unige.ch}
\affiliation{D\'epartement de Physique Th\'eorique and Center for Astroparticle Physics, Universit\'e de Gen\`eve, 24 quai Ansermet, CH--1211 Gen\`eve 4, Switzerland}

\date{\today}

\begin{abstract}
We present results from the first simulations of networks of Type I Abelian Higgs cosmic strings
 to include both matter and radiation eras and Cosmic Microwave Background (CMB) constraints.
In Type I strings,
the string tension is a slowly decreasing function of the ratio of the scalar and gauge mass-squared, $\beta$. 
 We find that the mean string separation shows no dependence on $\beta$, and that
the energy-momentum tensor correlators decrease approximately in proportion to the square of the string tension, with additional O(1) correction factors which asymptote to constants below $\beta \lesssim 0.01$. 
Strings in models with low self-couplings can therefore satisfy current CMB bounds at higher symmetry-breaking scales. This is particularly relevant for models where the gauge symmetry is broken in a supersymmetric flat direction, for which the effective self-coupling can be extremely small. 
If our results can be extrapolated to $\beta  \simeq 10^{-15}$, even strings formed at $10^{16}$ GeV (approximately the grand unification scale in supersymmetric extensions of the Standard Model) 
can be compatible with CMB constraints.
\end{abstract}


\maketitle


\section{Introduction} \label{sec:intro}

Topological defects are a common by-product of phase transitions 
in the early Universe \cite{Kibble:1976sj}.  
Many kinds of defect are possible \cite{VilShe94}, but perhaps the most well-motivated for study in cosmology are cosmic strings: they are expected for many realistic multi-field models of inflation \cite{Jeannerot:2003qv}, and they remain a constant small fraction of the energy density throughout the history of the Universe \cite{Zeldovich:1980gh,Vilenkin:1981iu}, neither dominating nor disappearing. 
More about the cosmic string scenario can be found in review articles 
\cite{Hindmarsh:1994re,Sakellariadou:2006qs,Copeland:2011dx,Hindmarsh:2011qj}.

If cosmic strings are formed then they will create characteristic observational signals from their 
metric perturbations and their decay products. The most significant are Cosmic Microwave Background (CMB) fluctuations \cite{Gott:1984ef,Kaiser:1984iv}, gravitational waves \cite{Vilenkin:1981bx,Hogan:1984is}, and high-energy cosmic and gamma rays \cite{Sigl:1995kk,Protheroe:1996pd}.  The relative proportion of the energy which goes into gravitational waves and high-energy particles is uncertain, as the short lifetime of loops of string in numerical simulations is contrary to naive expectation and not understood  \cite{Hindmarsh:2011qj} (See also \cite{Matsunami:2019fss} for recent simulations of decaying loops in the Abelian Higgs model).  CMB fluctuations are much less uncertain, and therefore provide the most robust constraints. 
The gravitational wave \cite{Ringeval:2017eww,Blanco-Pillado:2017rnf} and high-energy particle constraints \cite{Mota:2014uka} are to a large extent complementary.

CMB constraints have been studied in detail for the simplest kind of cosmic string, the Nielsen-Olesen vortex solution \cite{Nielsen:1973cs} 
of the Abelian Higgs model, for a particular set of scalar and gauge couplings \cite{Bevis:2007gh,Ade:2013xla,Lizarraga:2014xza,Ade:2015xua,Lizarraga:2016onn}, and in the Nambu-Goto approximation to the Nielsen-Olesen vortex network \cite{Ade:2013xla,Lazanu:2014eya,Ade:2015xua,Charnock:2016nzm}.
 
The CMB observations limit the contribution of cosmic strings to about 1\% in the temperature power spectrum, which pushes the string tension well below the grand unification scale. 
These limits start to rule out some scenarios, especially supersymmetric hybrid inflation models \cite{Battye:2010hg}, where the string tension and the amplitude of inflationary perturbations are both set by the ratio of the symmetry-breaking scale to the Planck scale. 

However, the string tension is reduced for a given symmetry-breaking scale if the scalar self-coupling is very small \cite{Hill:1987qx}, and the strings are Type I vortices. A similar effect occurs if the symmetry-breaking field takes an expectation value in a flat direction \cite{Lazarides:1986di,Perkins:1998re,Cui:2007js}, where the effective value of $\beta$ is given by a power of the ratio of the supersymmetric soft mass to the scalar vacuum expectation value.  The lower string tension can be used to construct models of supersymmetric hybrid inflation consistent with CMB limits \cite{Hindmarsh:2012wh}.

In more detail, the classical dynamics of the Abelian Higgs model are a function of $\beta = \lambda/2e^2$, where $\lambda$ is the scalar self-coupling and $e$ the gauge coupling.  
The string tension with complex scalar expectation value $\phi_0$ is 
$\mu(\be) = 2\pi \phi_0^2B(\be) $, where $B$ is a slowly-varying function of its argument.
All previous CMB limits are based on simulations with $\beta = 1$ (Bogomolnyi coupling \cite{Bog76}), for which 
$B(\be) = 1$.  The limit on the string tension $G\mu < 2.0 \times 10^{-7}$ can therefore be translated into a limit on the symmetry-breaking scale $\phi_0 < 2.2 \times 10^{15}$ GeV \cite{Lizarraga:2016onn}.

It is not clear how to extrapolate the CMB bounds obtained at $\be=1$. 
 One might have guessed that the string density was independent of $\beta$, and that 
one can simply scale the CMB signal by the appropriate power of the ratio of string tensions $\mu(\be)/\mu(1)$. However, 
changing the scalar self-coupling may change the energy loss rate, and hence the string density.  Furthermore, strings interact differently when $\beta<1$: parallel strings attract, and may form bound states whose flux is a multiple of $2\pi/e$.  The probability of string reconnection, which is important for energy loss from long strings, is also known to change with $\be$ \cite{Achucarro:2006es,Verbiest:2011kv}.
Moreover, the first network simulations of the Abelian Higgs model with $\be < 1$  in the radiation era \cite{Hiramatsu:2013tga} 
produced the surprising result that the string density peaked at around $\beta = 0.4$, and then decreased towards their minimum coupling ratio, $\beta = 0.2$.  

This then motivates the goals of this paper: to compute the scaling properties of Type I strings at lower $\beta$ than before, to characterise how the CMB anisotropies differ from those at Bogomolnyi coupling, and to extrapolate the results to even lower $\beta$.
We will see that our simulations,  which reach $\beta = 0.01$ and are four times larger and longer than those in \cite{Hiramatsu:2013tga}, provide good evidence that 
 the string density is indeed independent of $\beta$, and that the temperature
anisotropies can be scaled from those at $\beta=1$, but not simply by a power of the string tension ratio. Rather, there is an extra correction factor, which appears to converge to a constant for $\beta \lesssim 0.03$, whose value is estimated to be $0.52 \pm 0.03$.  We therefore can make predictions for very low $\beta$, supporting the argument that supersymmetric F-term hybrid inflation does not violate the CMB bounds due to strings.  We can also estimate what $\beta$ is required to render a given symmetry-breaking scale consistent with the constraints on strings. 

The outline of the paper is as follows: In the following section we introduce the physical model that we will use to simulate the formation and evolution of Type I strings. In Section \ref{sec:numerics} we describe the numerical approach used, and comment on the differences between the simulations of strings with and below the Bogolmonyi coupling. We present the simulation results in Section \ref{sec:results}, where we also show how these results can be scaled to the $\beta=1$ case. in Section \ref{sec:cls} we use the scaling results to translate the Bogolmonyi string CMB anisotropy power spectra $C_\ell$ and bounds to $\beta < 1$ and $\beta \ll1$.  In the final section we present our conclusions, including the implications for the symmetry-breaking scale.


\section{Abelian Higgs model and Type I strings} \label{sec:model}

The action for the Abelian Higgs (AH) model in a background metric $g_{\mu\nu}$ is 
\bea
S &=& -\int d^4x \, \sqrt{-g} \Big( g^{\mu\nu}D_\mu\phi^*D_\nu\phi +V(\phi) \nonumber\\
&& + \frac{1}{4e^2}g^{\mu\rho}g^{\nu\si}F_{\mu\nu}F_{\rho\si}\Big),
\eea
where $\phi(x)$ is a complex {scalar} field,  
$A_\mu(x)$ is a {vector} field, 
the covariant derivative is \(D_\mu = \partial_\mu - iA_\mu\), 
and the potential is \(V(\phi) = \frac{1}{4}\la(|\phi|^2 - \phi_0^2)^2\).

By rescaling fields and coordinates
\ben
\label{e:Rescale}
\tilde{x}^\mu = e\phi_0 x^\mu, \quad \tilde{A}_\mu = A_\mu/e\phi_0, \quad \tilde\phi = \phi/\phi_0,
\een
the action becomes
\bea
S &=& - \frac{1}{e^2}\int d^4\tilde{x} \, \sqrt{-g} \Big( g^{\mu\nu}\tilde{D}_\mu\tilde{\phi}^*\tilde{D}_\nu\tilde{\phi} +\tilde{V}(\tilde{\phi}) \nonumber\\
&& + \frac{1}{4}g^{\mu\rho}g^{\nu\si}\tilde{F}_{\mu\nu}\tilde{F}_{\rho\si}\Big),
\eea
where the dimensionless potential is 
\ben
\tilde{V}(\tilde{\phi}) = \half \be(|\tilde{\phi}|^2 - 1)^2,
\een
with 
\ben
\be = {\la}/{2e^2}.
\een
Through the rescaling, it becomes clear that $\be$ is the sole free parameter of the classical theory.

Specialising to Minkowski space-time, 
the field equations have static cylindrically symmetric solutions, Nielsen-Olesen (NO) vortices \cite{Nielsen:1973cs},
which can be written 
\ben
\phi =  f(\rtil)e^{i\th}, \quad A_i = \hat\varphi_i {g(\rtil)}/{\rtil}, 
\label{e:NOAns}
\een
where $\rtil = e\phi_0 r$ is a dimensionless radial cylindrical coordinate, and $\varphi$ the angular coordinate.
Finite energy per unit length is obtained if $f \to 1$ and $g \to 0$ as $\rtil \to \infty$.  
Regularity requires that $f(0) = 0$ and $g(0) = 0$.
This solution represents a string of energy-momentum arranged along the $z$ axis.
At the core of the string there is a magnetic field 
\ben
B_i = \hat{z}_i  {g'(\rtil)}/{\rtil}.
\een
The functions $f$ and $g$ approach the vacuum as
\bea
f &\simeq &
          1-f_1 \rtil^{-1/2}\exp(-\sqrt{\beta} \rtil), \\
g &\simeq& 1-g_1\rtil^{1/2}\exp(-\rtil), 
\label{e:NOAsy}
\eea
where $f_1$ and $g_1$ are constants. 
Hence the gauge field approaches the vacuum over the physical distance scale $\SwidG = (e\phi_0)^{-1}$, 
and the scalar field over a length  $\SwidS = (\sqrt{\la/2}\phi_0)^{-1}$. 
Note that when $\be = 1$ these distance scales are equal.
It can be shown that 
\ben
\label{e:StrTen}
\mu = 2\pi\phi_0^2 B(\be)\,,
\een
with $B(1) = 1$ \cite{Bog76,BogVai76,Manton:2004tk}.
Away from $\be = 1$, the function can be established only numerically, and 
has the following approximate behaviour\footnote{We obtained the fit  for the range $10^{-6} \lesssim \be \lesssim 10^2$ from the digitised graph of $B(\beta)$ against $2/\beta$ in Fig.~1b of reference \cite{Hill:1987qx}, with a maximum error of 7\% error over the range.}
\bea
\label{e:StrTenFun}
\log{B(\be)} &\simeq& \sum_{n=1}^3 \Lambda_n \log^n(\be), \quad\quad\quad  10^{-6} \lesssim \be \lesssim 10^2, \nonumber \\
B(\be) &\simeq& 2.4/\ln(2/\be), \quad\quad\quad \be \lesssim 10^{-6}\,,
\eea
where $\Lambda_1 = 0.195$, $\Lambda_2 = 0.013$ and $\Lambda_3 = 0.0004$.

For $\be< 1$ there is a short-range attractive force between parallel strings, while the force is repulsive for $\be >1$.  At $\be = 1$ the force vanishes \cite{Manton:2004tk}. One can interpret this behaviour as there being an attractive force from the scalar field with range $\SwidS$, and a repulsive gauge force with range $\SwidG$.

In the cosmological context, 
the relevant metric is the spatially flat Robertson-Walker (FLRW) metric  
\ben
g_{\mu\nu} = a^2(\ta)\eta_{\mu\nu}
\een
where $\eta_{\mu\nu} = \text{diag}(-1,1,1,1)_{\mu\nu}$ is the Minkowski metric, 
$\ta$ is conformal time, and $a(\ta)$ is the scale factor.
In the radiation era $a(\ta) \propto \tau$, and in the matter era \(a \propto \ta^{2}\).

The resulting field equations in the temporal gauge ($A_0 = 0$) are 
\bea
\ddot\phi + 2\frac{\dot a }{a}\dot \phi -\bD^2\phi + a^2 \la (|\phi|^2 -1)\phi &=& 0, \\
\pa^\mu\left( \frac{1}{e^2} F_{\mu\nu}\right)  - ia^2(\phi^*D_\nu\phi - D_\nu\phi^*\phi) &=& 0,
\eea
where indices are raised with the Minkowski metric. In an expanding universe, the Minkowski space straight string solutions remain a very good approximation, as corrections are of order  $H\SwidS$ and $H\SwidG$, where $H = \dot a /a$ is the Hubble rate in conformal time.

The physical length scales $\SwidS$, $\SwidG$
shrink in comoving coordinates, so a numerical solution on a fixed comoving grid the string core must start off large in order to remain resolved thoughout the simulation.  

There is a potential problem with the string core being larger than the Hubble radius at early times, which can be avoided by modifying the equations so that the comoving  width of the core grows in the early phase of the simulations \cite{Press:1989yh,Bevis:2006mj}. This can be accomplished by modifying the equations to read 
\bea
\ddot\phi + 2\frac{\dot a }{a}\dot \phi -\bD^2\phi + a^{2s_\phi} \la (|\phi|^2 - 1)\phi &=& 0, \label{eq:EoM_1}\\
\pa^\mu\left(\frac{{a^{2(1-s_A)}}}{e^2} F_{\mu\nu} \right) - ia^2(\phi^*D_\nu\phi - D_\nu\phi^*\phi) &=& 0. \label{eq:EoM_2}
\eea
One can view this change as the introduction of time-dependent couplings 
 \ben
  e^2(\ta) = e^2_0a^{2(s_A-1)}(\ta), \quad \la(\ta) = \la_0a^{2(s_\phi-1)}(\ta),
\een
while keeping the vacuum expectation value $\phi_0$ fixed.

When $s=0$, the width of the regions where the fields depart from their vacuum values is fixed in comoving coordinates, while negative $s_{\phi,A}$ causes the widths of the scalar and gauge cores to grow in comoving coordinates. 
The true dynamics are recovered at $s=1$, which is adopted for the bulk of our simulations.
 
In the next section we shall show how this system of equations is solved numerically, how the initial conditions are prepared, and which quantities we measure.


\section{Numerical methods and measurements} \label{sec:numerics}

\subsection{Methods}

Full details of the numerical method used to discretise the equations of motion (\ref{eq:EoM_1}) and (\ref{eq:EoM_2}) can be found in \cite{Bevis:2007qz,Hindmarsh:2017qff}. 
Here we highlight differences with those simulations.

For the initial conditions, at time $\tStart$ we set the scalar field $\phi$ to be a static gaussian random field with $\vev{|\phi^2|} = \phi_0^2$ and correlation length $\corrLen$ around $5\times$ the width of the scalar core (see Table~\ref{table:param}). A cooling period is used at the beginning of the simulation, between $\tStart$ and $\tDiff$, to smooth the field distribution by applying a diffusive evolution 
\bea
\dot{\phi} &=& D_jD_j\phi - \frac{\lambda}{2}(|\phi|^2 - \phi_0^2)\phi \, ,
\label{AHdiff1}\\
F_{0j} &=& \partial_iF_{ij} - e^2 {\rm Im} (\phi^*D_j\phi) \, .
\label{AHdiff2}
\eea
The values of $\la$ and $e$ during this phase are both O(1), in order to accelerate the relaxation of the scalar field to the potential minimum, and the generation of flux in the string cores. 

 This cooling phase is not present in the simulations of Ref.~\cite{Hiramatsu:2013tga}, which renders the observables more sensitive to the thermal excitations present in the initial state.

After the cooling phase,  we allow the string cores to grow by setting  $s_{\phi,A}$ negative, so that they meet their true comoving widths at time time $\tCG$. As the scalar core is larger than the gauge core for $\beta <1$, we must have $|s_{\phi}| > |s_A|$. 

Finally the true evolution of the network dictated by Eq.~(\ref{eq:EoM_1}) and (\ref{eq:EoM_2}) is recovered until the end of the simulation, $\tEnd$.

Table \ref{table:param} shows the parameters used in the simulations performed. 
Where they are not specified, they are the same as in  \cite{Hindmarsh:2017qff}. 
In particular, the lattices contained $4096^3$ points, with comoving spatial separation $dx=0.5\phi_0^{-1}$ and time steps of $dt=0.2dx$. 
One run was performed for each set of parameters listed in the table.

\begin{table}[h!]
\renewcommand{\arraystretch}{1.2}
\begin{tabular}{|c||c|c|c|c||c|c|}
\hline
Cosmology & \multicolumn{4}{c||}{Radiation} &  \multicolumn{2}{c|}{Matter} \\\hline
$\be$ & 0.01 & 0.025 & 0.1 & 0.25 & 0.01 & 0.025 \\\hline
$e_0$ & 1 & 1 & 1 & 1 & 1 & 1 \\
$\lambda_0$ & 0.02 & 0.05 & 0.2 & 0.5 & 0.02 & 0.05 \\
$\corrLen$ & 50 & 32 & 16 & 10 & 50 & 31 \\
$\tStart$ & -26 & -23.6 & -17 & -10 & -21 & -16 \\
$\tDiff$ & 4 & 6.4 & 13 & 20 & 9 & 14 \\
$-s_\phi$ & 1.0  & 1.0  & 1.0  & 1.0  & 0.5  & 0.5  \\
$-s_A$ & 0.423	& 0.48	& 0.6	 & 0.708 & 0.218&	0.245\\
$\tCG$ & 220 & 220 & 220 & 200 & 500 & 500 \\
$\tRef$ & 450 & 450 & 450 & 450 & 600 & 600 \\
$\tEnd$ & 1200 & 1200 & 1200 & 1200 & 1200 & 1200\\\hline

\end{tabular}
 \caption{\label{table:param} Run parameters. See text for explanation.   }
\end{table}

\subsection{Measurements}
The main indicator of the length scale of the string network is the 
mean string separation $\xi$, defined as
\begin{equation}
\xi = \sqrt{\frac{V}{\Slen}},
\label{eq:xi}
\end{equation}
where $V$ is the comoving volume and $\Slen$ the total length of string. 
The mean string separation is therefore the inverse square root of the length density.

In this work we estimate $\Slen$ by counting the number  of plaquettes pierced by string, as measured by non-zero gauge-invariant winding number around the boundary \cite{Kajantie:1998bg}. The resulting length is corrected by the factor $\pi / 6$ as proposed in \cite{Scherrer:1997sq} to reduce the Manhattan effect of the cubic mesh.
We denote the resulting string separation estimate by  $\xi_{\text{w}}$.

In previous work we have also used the mean Lagrangian to estimate the string length, using the fact that the Lagrangian is proportional to the invariant length of a Nambu-Goto string, or 
\begin{equation}
\Slen = \mathcal{\bar{L}} V / \mu,
\label{eq:l_lag}
\end{equation}
with $\mathcal{\bar{L}}$ being the mean Lagrangian density and $\mu$ the string tension.
The resulting estimator is denoted $\xi_\mathcal{L}$.

The Lagrangian estimator assumes that the any oscillatory modes in the field are perturbative, for which the Lagrangian density vanishes.  In the ideal case the estimators should be proportional to one another.
However, we will see that the Lagrangian estimator is not a good one for the lowest values of $\beta$ for which we run, showing oscillatory behaviour. 
We trace the problem back to entering the core growth phase too early, at a time when the string width increases at too high a fraction of the Hubble rate (0.5 for $\beta = 0.01$). This triggers homogeneous oscillations in the field.

The key observables for CMB predictions are the unequal time correlators of the energy-momentum tensor (UETCs), which are defined as follows:
\begin{equation}
{U}_{\lambda\kappa\mu\nu} (\textbf{k},\tau,\tau') = \langle{\mathcal{T}}_{\lambda\kappa}(\textbf{k},\tau){\mathcal{T}}^*_{\mu\nu}(\textbf{k},\tau')\rangle\,,
\label{emtens}
\end{equation}
where ${\mathcal{T}}_{\alpha\beta}(\textbf{k},\tau)$ is the AH energy-momentum tensor. 

By rotational symmetry, and  for linearised cosmological perturbations, the problem separates into (decoupled) scalar, vector and tensor correlators. Both vectors and tensors have two components, which are related by parity in the AH model, so therefore there is only one vector and one tensor independent UETC. The scalars have three different independent UETC. Thus, the problem reduces to 5 independent correlators that depend on 3 variables : $k$ (the magnitude of $\bk$), $\tau$ and $\tau'$.  Actually, when the network is scaling, the UETCs depend only on 2 variables via the combinations $k\tau$ and $k\tau'$.

Thus we can write the scaling UETCs as 
\begin{equation}
{U}_{ab}(\textbf{k} ,\tau,\tau') = \frac{\phi_0^4}{\sqrt{\tau\tau'}}\frac{1}{V}{C}_{ab}(k\tau,k\tau'),
\label{UETCdecom}
\end{equation}
where $\phi_0$ is the symmetry breaking scale, $V$ a formal comoving volume factor. The indices $a$, $b$ indicate the independent components of the energy momentum tensor: two scalar, one vector and one tensor. We will denote the scalar indices $1$ and $2$ (corresponding to the longitudinal gauge potentials $\phi$ and $\psi$), the vector component with `v' and the tensor component with `t'.

The full information encoded in the UETCs can be better analysed using two related functions: the (scaling) equal time correlators (ETC) and the decoherence functions. The ETCs are defined as,

\ben
E_{ab}(k\ta) = C_{ab}(k\ta,k\ta).
\label{ETC}
\een 
These are numerically cheaper to obtain, but are a good measure for the scaling of the network (see next section) and they will allow us to get all the necessary information to predict the CMB anisotropies. 

The decoherence functions exhibit the temporal decoherence properties of the UETCs. They are defined as

\ben
D_{ab}(k\ta,k\ta') = \frac{C_{ab}(k\ta,k\ta')}{\sqrt{E_{aa}(k\ta)E_{bb}(k\ta')}}.
\een

These functions, except the cross-correlator decoherence function, 
are symmetric under $k\ta \leftrightarrow k\ta'$ and equal to 1 at equal-times. At equal times $D_{12}$ measures the~(anti)correlation~between the two scalar sources; in general is not equal to $1$, and is in fact negative.

We compute the UETCs at 150 times $\tau$ between $\tRef$ and $\tEnd$, by cross-correlating the projected components of the energy momentum tensor described above with those computed at time $\tRef$.


\section{Results} \label{sec:results}

The basic requirement for extrapolating the results of a numerical simulation to cosmological scales is scaling, by which we mean that network observables with the dimensions of (comoving) length increase with (conformal) time. We show  evidence  that our simulations have reached a scaling regime by means of different measures of the mean string separation $\xi$, and also by the study of equal time correlators (ETCs) \cite{Bevis:2010gj,Daverio:2015nva}.

\begin{figure}
\centering
\includegraphics[width=0.5\textwidth]{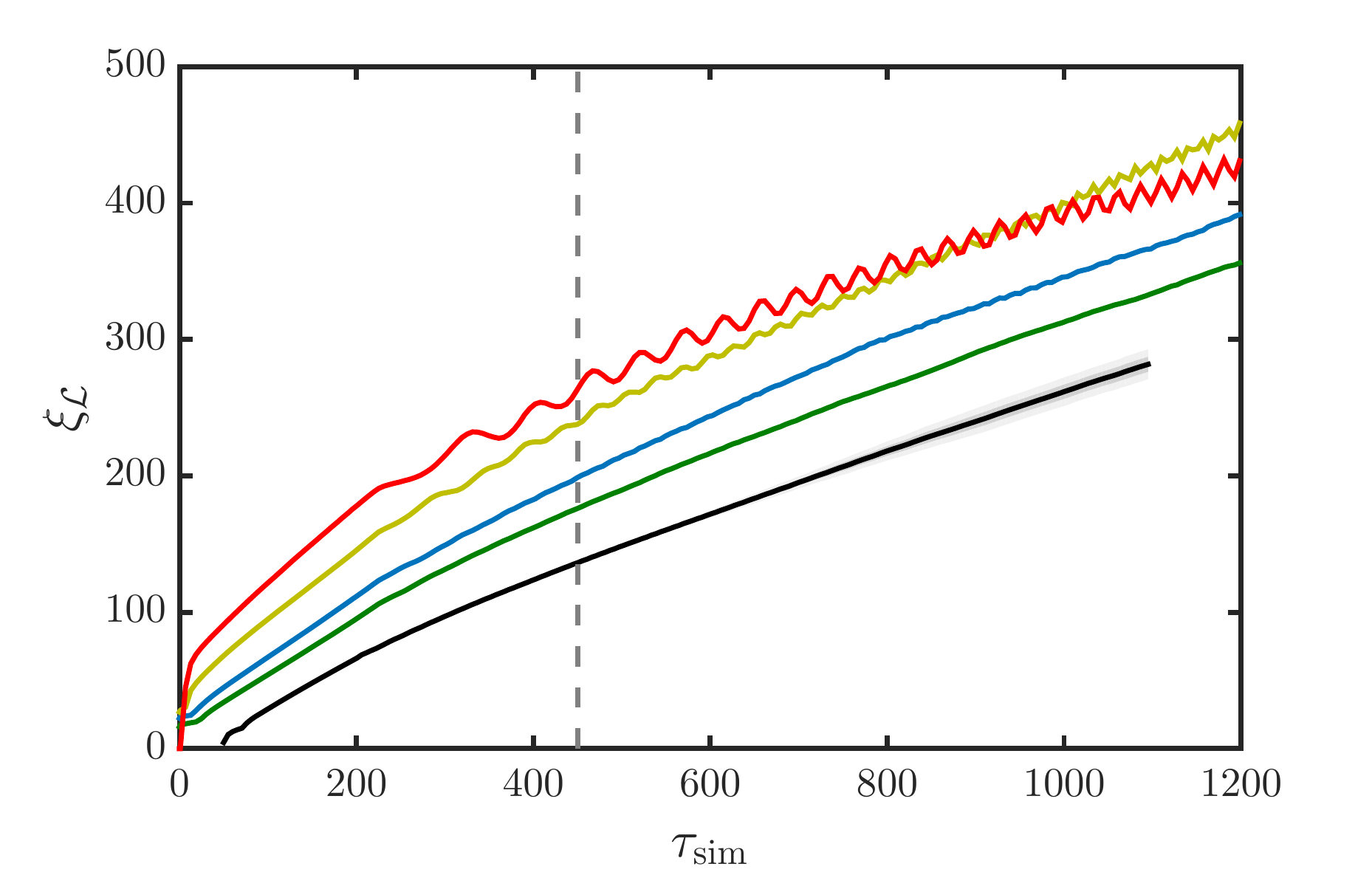}\\
\includegraphics[width=0.5\textwidth]{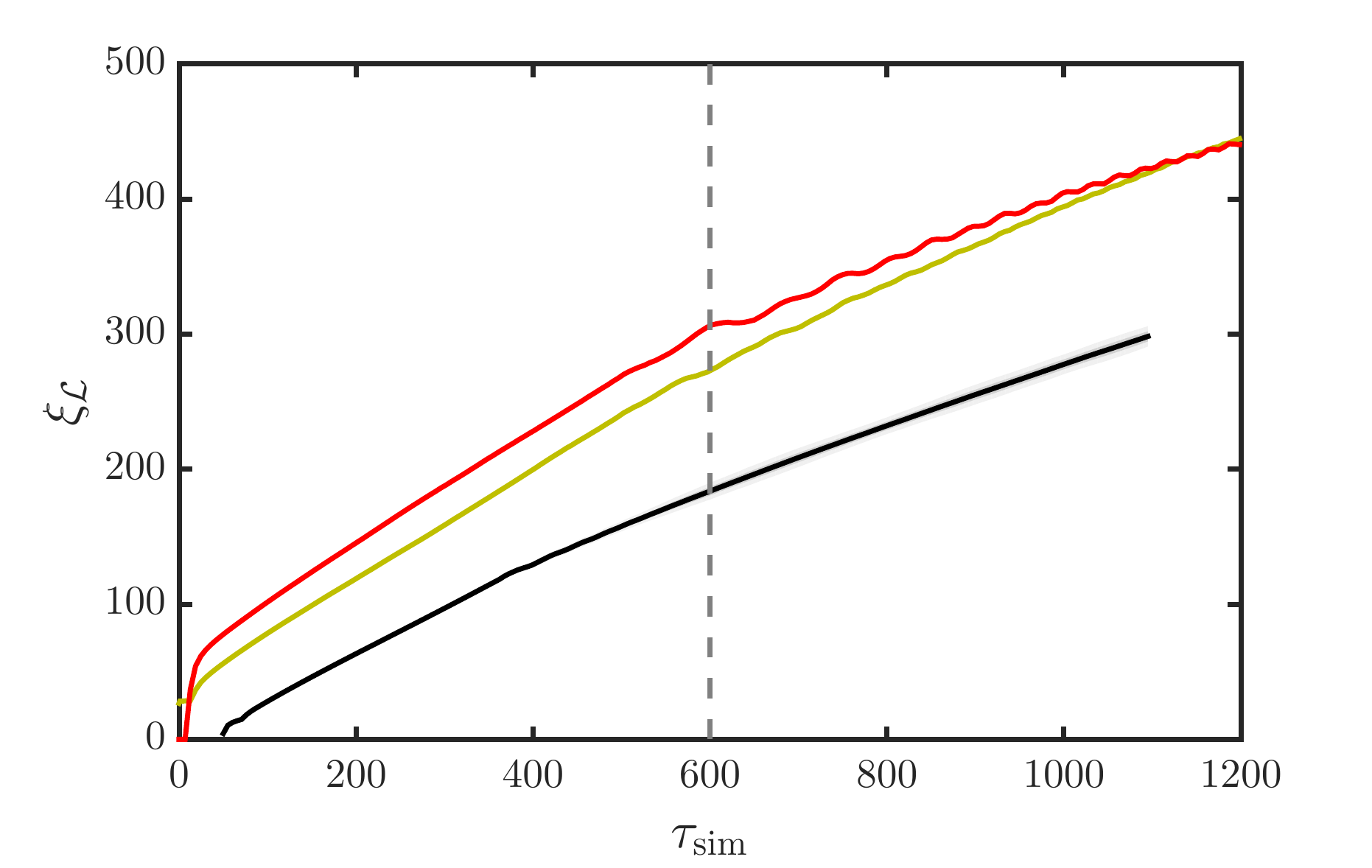}
\caption{Figures showing mean string separation $\xi$ for radiation (top)  and matter (bottom) eras, using a Lagrangian weighted measures. In radiation, the different lines correspond, from top to bottom, to values of $\beta=0.01\,,0.025\,,0.1\,,0.25$ and $1.0$ respectively. In matter, the lines correspond to  $\beta=0.01\,,0.025$ and $1.0$ respectively. The vertical line corresponds to $\tRef$.
\label{Fig:xilag} }
\end{figure}

\begin{figure}
\centering
\includegraphics[width=0.5\textwidth]{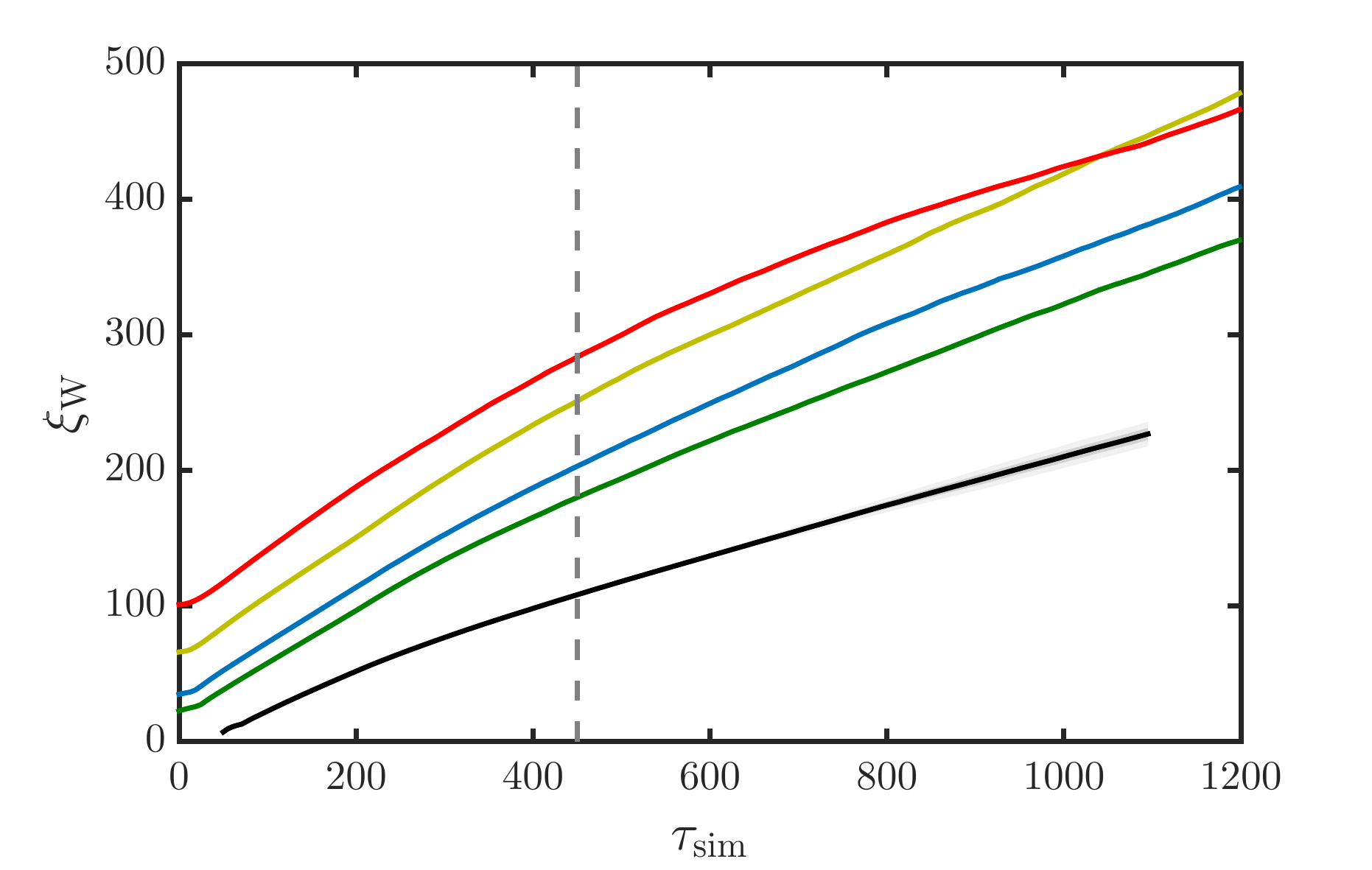}\\
\includegraphics[width=0.5\textwidth]{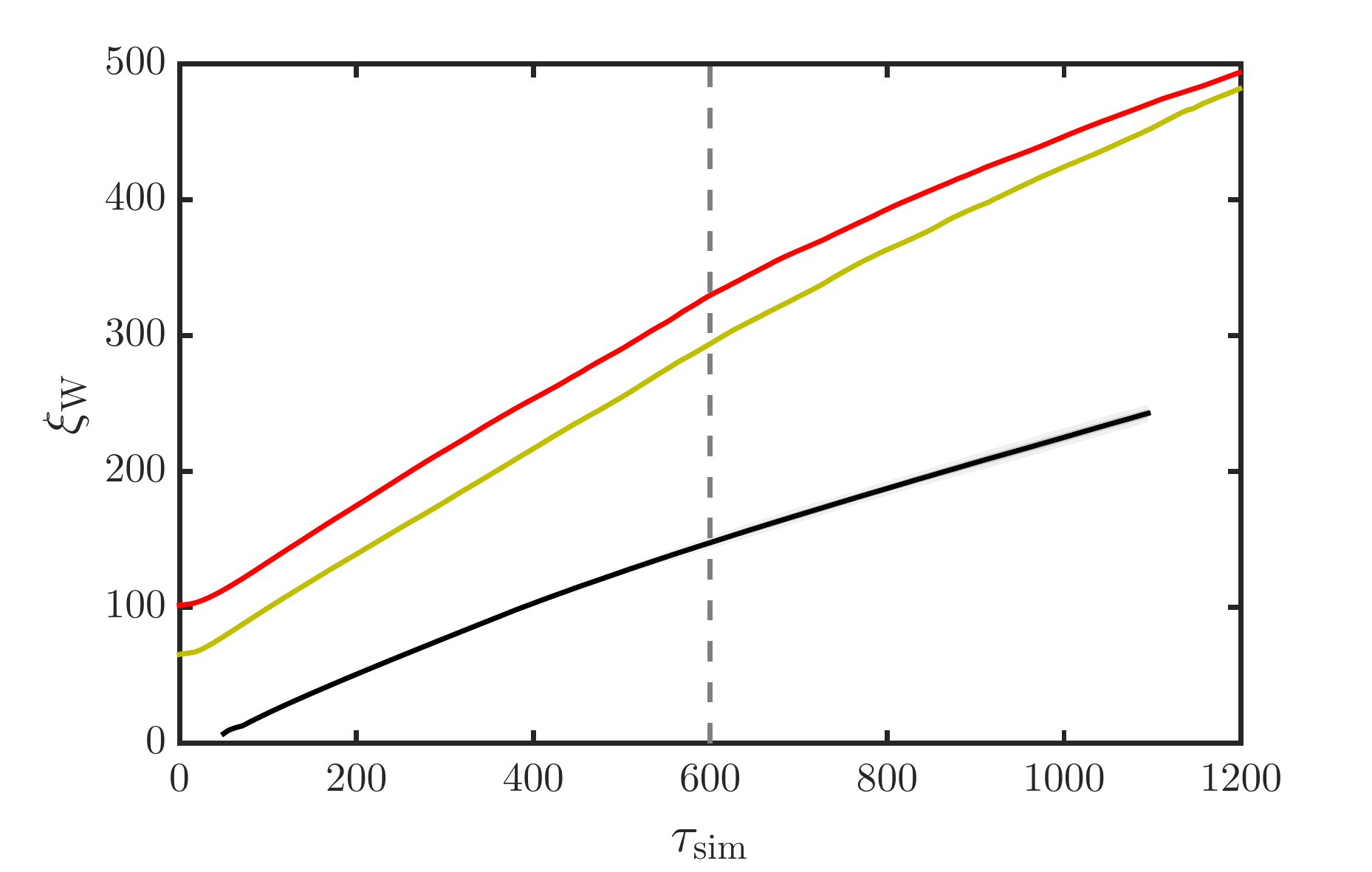}
\caption{Figures showing the mean string separation $\xi$ for radiation (top) and matter (bottom), using the winding estimate. The colour scheme is the same as in Fig.~\ref{Fig:xilag}. The vertical line corresponds to $\tRef$.
\label{Fig:xiwinding}}
\end{figure}

Figures~\ref{Fig:xilag} and \ref{Fig:xiwinding} show the mean string separation estimated with two different measures: the total Lagrangian $\xi_\mathcal{L}$  and the winding measure $\xi_{\text{w}}$ respectively (see Eqs.~(\ref{eq:xi}) and (\ref{eq:l_lag})). Both figures include results from simulations in radiation dominated (top) and matter dominated backgrounds (bottom). The colour scheme is the following: red corresponds to $\beta=0.01$, yellow to $\beta=0.025$, blue to $\beta=0.1$, and green to $\beta=0.25$. As a reference $\beta=1$ from \cite{Daverio:2015nva} is also added in black.  In the figures we show a vertical line showing $\tRef$. Times earlier than this are not used in the study of the network, since they are just an artifact of the initial conditions.The evidence of scaling happens for times later than  $\tRef$. 
The figures  provide some evidence  that all simulations are reaching scaling, as towards the end of the simulations the behaviour of $\xi$ is consistent with a linear increase with conformal time with similar slopes. 

There are two phenomena worth mentioning: on the one hand, the mean string separation exhibits a late time attenuation for the lowest $\beta$, \ie\ , $\beta=0.01$, visible in both radiation and matter, and in both different $\xi$ measures (as is evident in the curves intersecting ). On the other, the Lagrangian measure is oscillatory for $\beta=0.01$ (and to a lesser degree for $\beta=0.025$).  The attenuation and the oscillations are caused by a poor choice of $\tCG$ in the initial conditions (see discussion of Sec.~\ref{sec:numerics}) which resulted in the scalar field executing coherent oscillations around its minimum. 
 
However, visual inspection of the measures and comparison between $\xiLag$ and $\xiWind$ show that the description of the evolution of the scaling regime can be equivalently described by both estimators.

\begin{figure}[h!]
\centering
\includegraphics[width=0.5\textwidth]{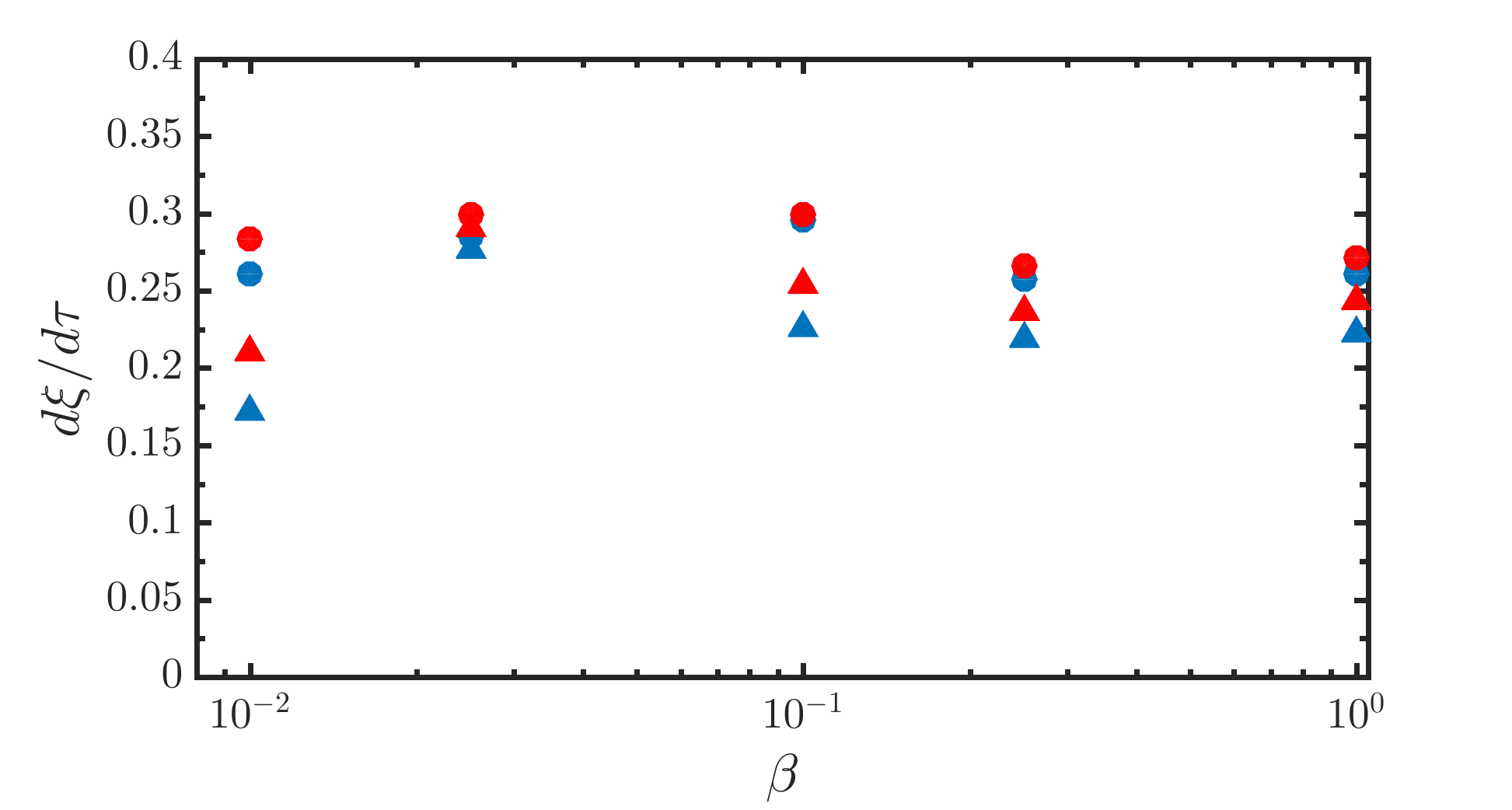}\\
\includegraphics[width=0.5\textwidth]{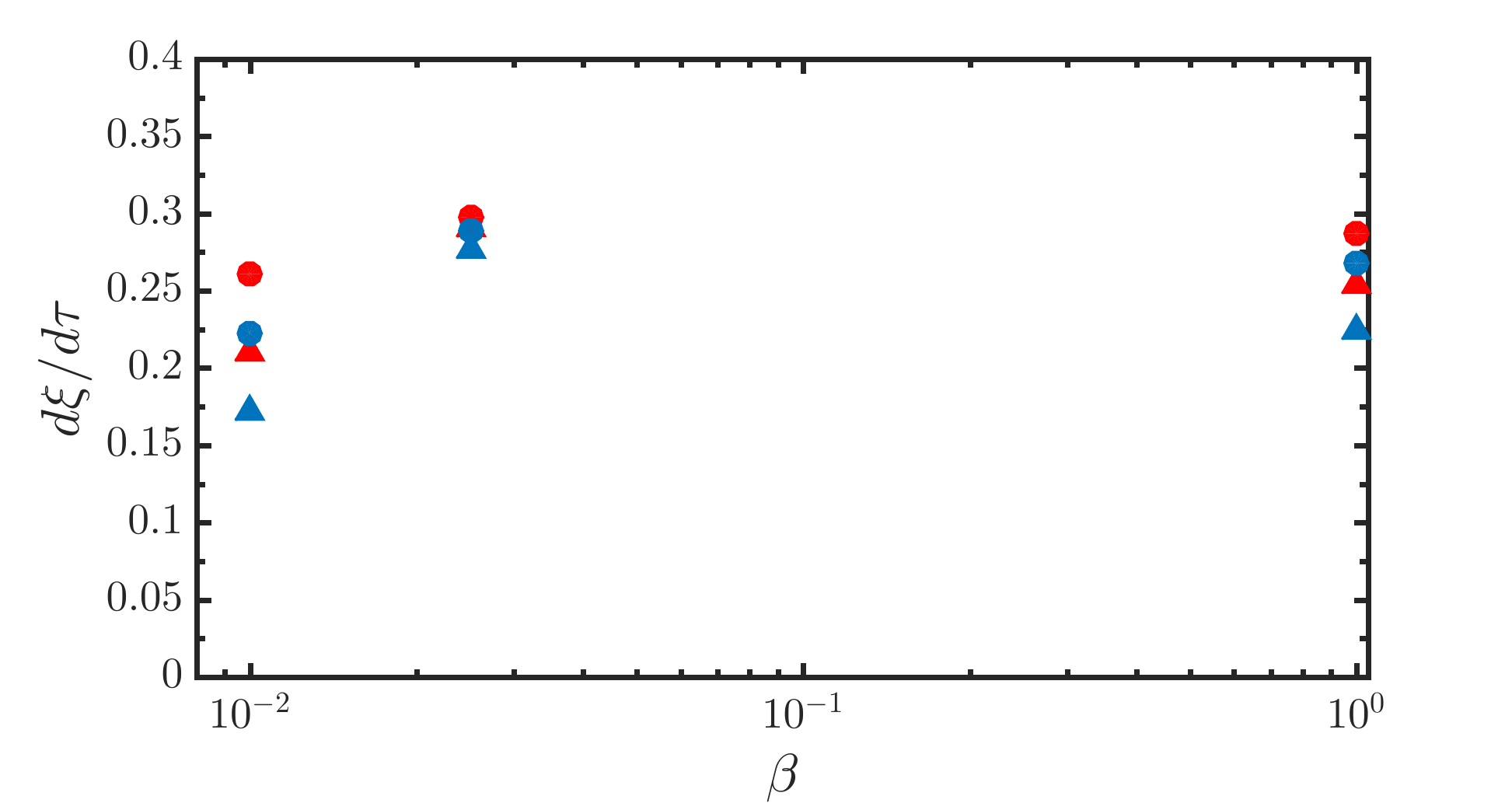}
\caption{ The rate of change with conformal time of the comoving mean string separation $\dot\xi$ for radiation and matter, using a winding (red) and Lagrangian (blue) weighted measures. The averages are calculated over times between 600 and 800 (depicted with a circles) and  between 1000 and 1200 (depicted with triangles).
\label{Fig:xidot}}
\end{figure}

This is also confirmed by checking the derivative of $\xi$ with respect to time. In Fig~\ref{Fig:xidot} we show  the values obtained from the Lagrangian (blue) and winding (red) estimators. In order to account for the time variation of the slope, we calculate the average value of $\dot{\xi}$ over two different time ranges: intermediate times between 600 and 800 (depicted with a circles), and late times at the range between 1000 and 1200 (depicted with triangles), all in simulation units. This selection encompasses a sufficiently large time range so that  the oscillations are averaged out. Fig~\ref{Fig:xidot} shows that the values given by both estimators are consistent and equivalent (and also shows  the late time attenuation exhibited by $\beta=0.01$). 

\begin{table}[h!]
\resizebox{0.48\textwidth}{!}{
\renewcommand{\arraystretch}{1.2}
\begin{tabular}{|c||c|c|c|c||c|c|}
\hline
Cosmology & \multicolumn{4}{c||}{Radiation} &  \multicolumn{2}{c|}{Matter} \\\hline
$\be$ & 0.01 & 0.025 & 0.1 & 0.25 & 0.01 & 0.025 \\\hline
$\tau_{\rm offset}$ & -747.78 & -387.01 & -228.51 & -238.16 & -495.00 & -284.52 \\\hline

\end{tabular}}
 \caption{\label{table:toffset} Values of $\tau_{\rm offset}$ for different simulations. All values obtained fitting $\xi$ over $\tSim \in [600\ 800]$.}
\end{table}

Another measure of scaling can be obtained by the ratio of mean string separation to time.  However, it is worth mentioning that in each simulation, 
the asymptotic linear behaviour has a time offset ($\tau_{\rm offset}$) due to the initial string density being somewhat lower than the scaling value, see Table~\ref{table:toffset} for specific values. 
(see \cite{Bevis:2010gj} for a more detailed discussion). One could consider that the string network is scaling with respect to an internal time kept by the network itself, rather than $\tSim$, the time recorded in the simulation.
The offset is larger for lower values of $\beta$ due to the higher initial string separation in these units.
This internal time of the simulation can be estimated from the slopes of the mean string separation $\alpha=d\xi/d\tau$ (see \cite{Daverio:2015nva}) according to  
\ben
\tau = \tSim-\tau_{\rm offset} = \al^{-1}(\tSim) \xi(\tSim)\,.
\label{eq:tau}
\een
Although in previous papers we used the Lagrangian estimator for this purpose,\footnote{We also made the unfortunate choice of using $\beta$ for the slope, which here conflicts with the coupling constant ratio.} in this case we chose the winding estimator to avoid possible problems caused by the oscillations of the Lagrangian estimator.  

\begin{figure}
\centering
\includegraphics[width=0.5\textwidth]{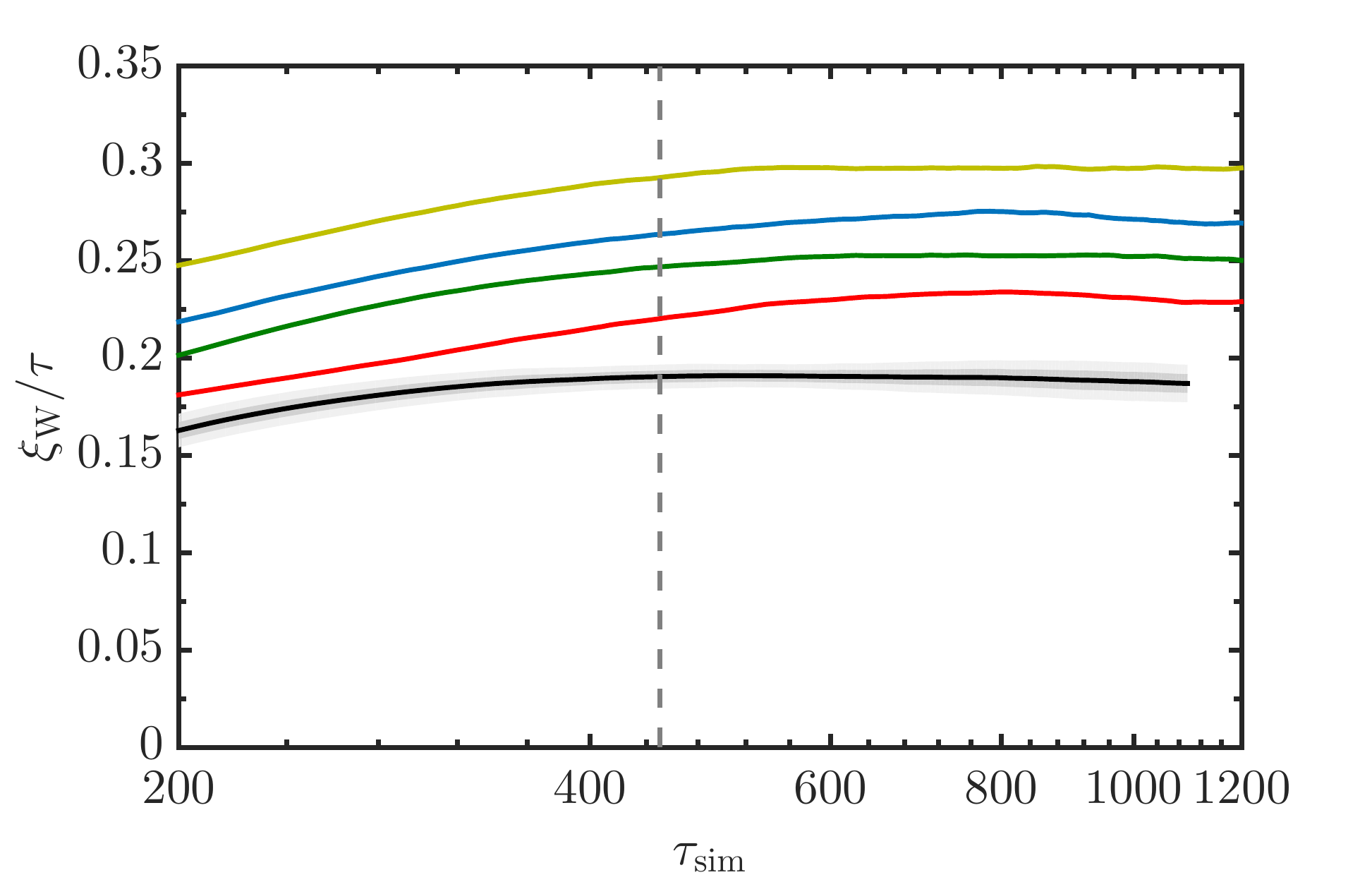}\\
\includegraphics[width=0.5\textwidth]{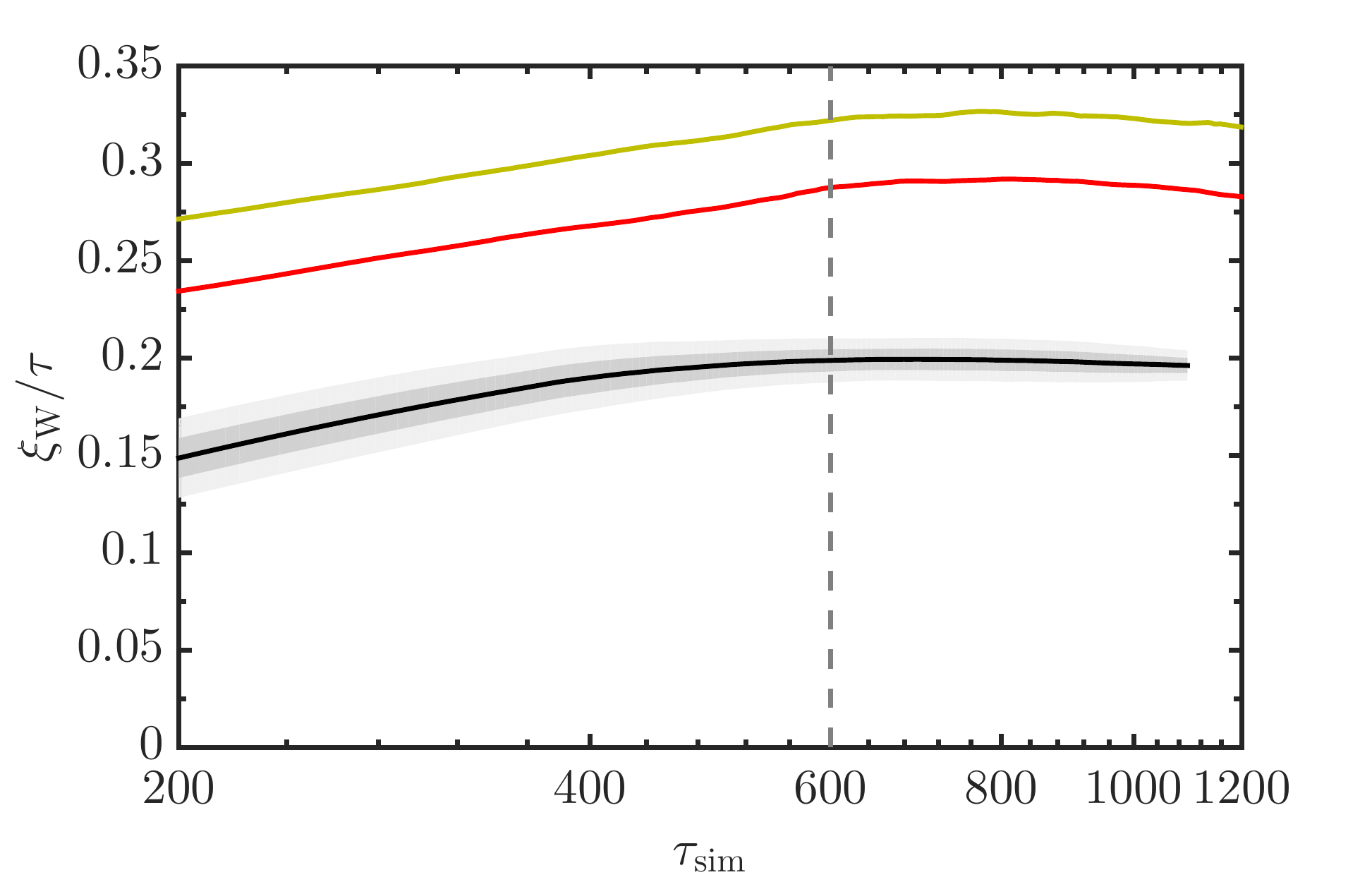}
\caption{Figures showing the ratio of the mean string separation to time for radiation (top)  and matter (bottom) eras, using the winding estimate. The colour scheme is the same as in previous similar plots.
\label{Fig:xiW_tau} }
\end{figure}

Using the winding estimator for the mean string separation and the internal time of the simulation,  in Fig.~\ref{Fig:xiW_tau} we show the ratio of the mean string separation to time ($\xiWind/\tau$), giving another indication of scaling.

The ETCs of the energy-momentum tensor (\ref{ETC}) can also be used to estimate scaling.  If the network scales, the ETCs from different stages of the simulation should collapse to a single line when plotted against $k\tau$.  Using the definition of time given in (\ref{eq:tau}), the $\xi$-scaled scaling ETCs read\footnote{In Eqs.~(21) and (28) of Ref.~\cite{Daverio:2015nva} this normalisation was mistakenly expressed using the inverse factor.}

\begin{equation}
E_{a}(k\tau,\beta)= \left(\frac{\tau}{\tSim}\right) E_{a}^{\rm (sim)}(k\tSim,\beta),
\end{equation}
where $\tau$ is the time as defined in Eq.~(\ref{eq:tau}), with $\xi$ and $\alpha$ determined from the winding measure. We represent simulation quantities with the subscript and superscript `sim'.

\begin{figure*}
\centering
\includegraphics[width=0.49\textwidth]{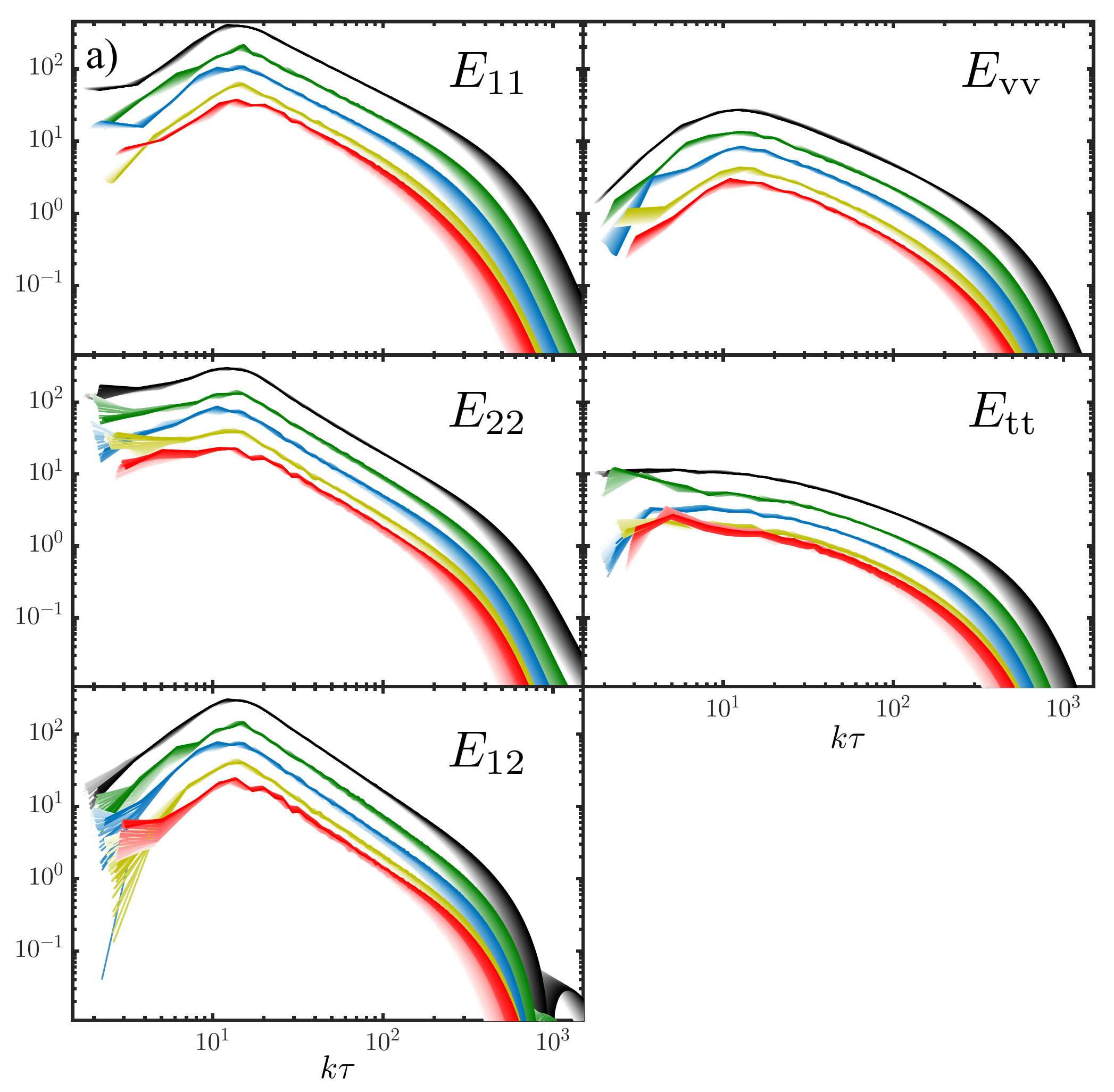}
\includegraphics[ width=0.49\textwidth]{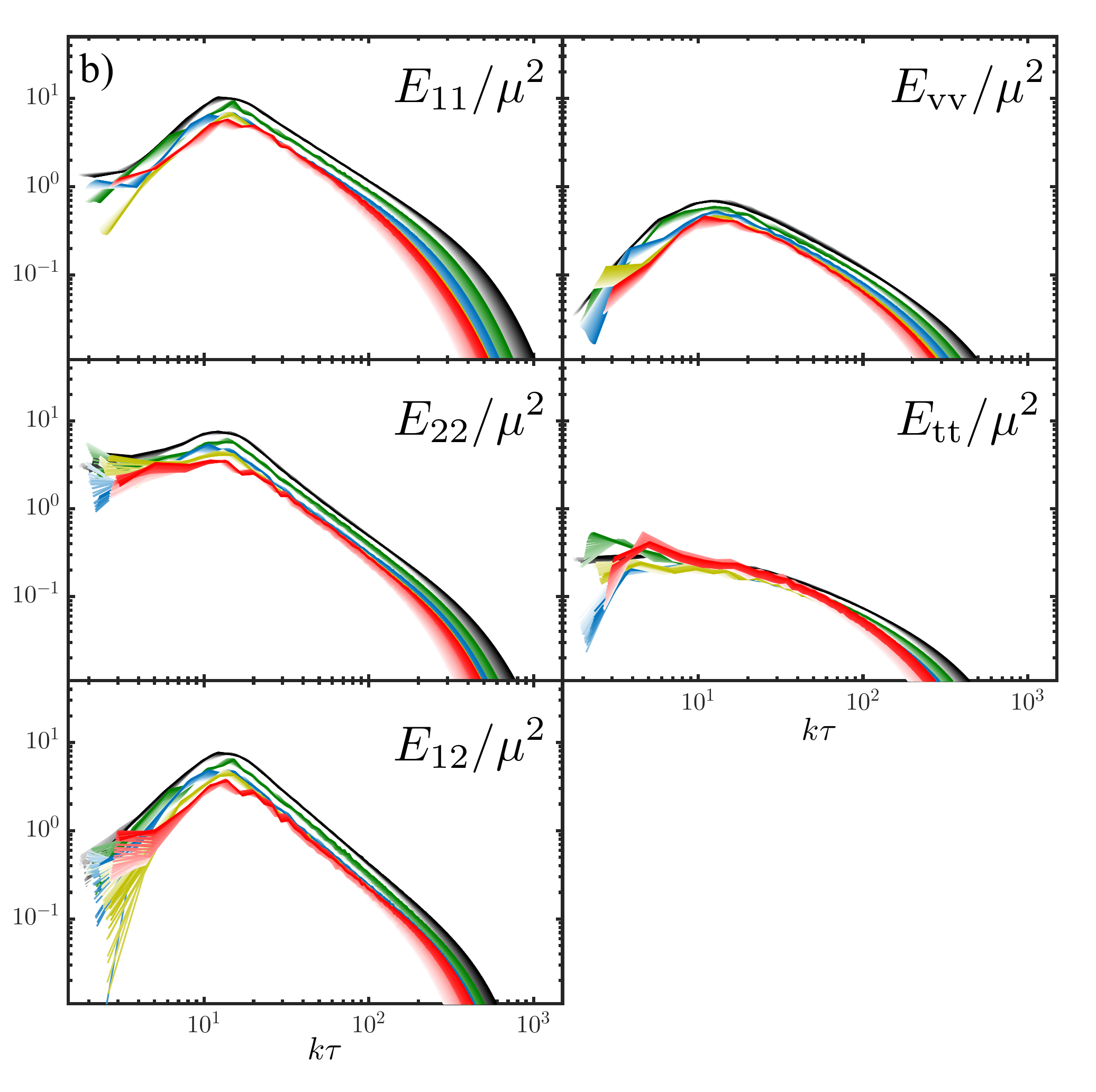}
\caption{Complete set of ETCs from simulations in the radiation era. There are two sets of panels (a and b), each containing five different ETCs:   on he left $E_{11}$, $E_{22}$ and $E_{12}$  from top to bottom, and on the right $E_{\text{vv}}$ and $E_{\text{tt}}$ from top to bottom. The a) panels show ETCs at their original amplitude, whereas the b) panels shows ETCs rescaled with the $\mu^2$ that corresponds to each $\beta$. The color scheme is as before: red for $\beta=0.01$, yellow for $\beta=0.025$, blue for $\beta=0.1$ green for $\beta=0.25$ and black for $\beta=1.0$. ETCs are extracted from $\tSim\in[600, 700]$ for $\beta<1$ and $\tSim\in[450, 600]$ for $\beta=1$. Within one colour, lighter shades correspond to earlier times. 
\label{Fig:ETCrad}}
\end{figure*}

The complete set of ETCs for different $\beta$s can be found in panel ``a)" of Fig.~\ref{Fig:ETCrad} for the radiation era. We also include the decoherence functions in Fig.~\ref{Fig:CohFunrad} from radiation domination (see Appendix for the corresponding matter era figures). We maintain the same colour scheme as in previous figures.  

The ETCs  also provide evidence  that for each case scaling has been reached. The shapes of the ETC for different values of $\beta$ seem similar, though lower values of $\beta$ correspond to lower values of the amplitude of the ETCs. Moreover, the differences in the values of the amplitudes are apparently shared between different ETCs. This difference could come from the different tension of the strings $\mu^2$. The decoherence functions also support this resemblance of the correlators of different simulations, indicating that the temporal decoherence of the UETCs is independent of $\beta$.

\begin{figure}
\centering
\includegraphics[width=0.49 \textwidth]{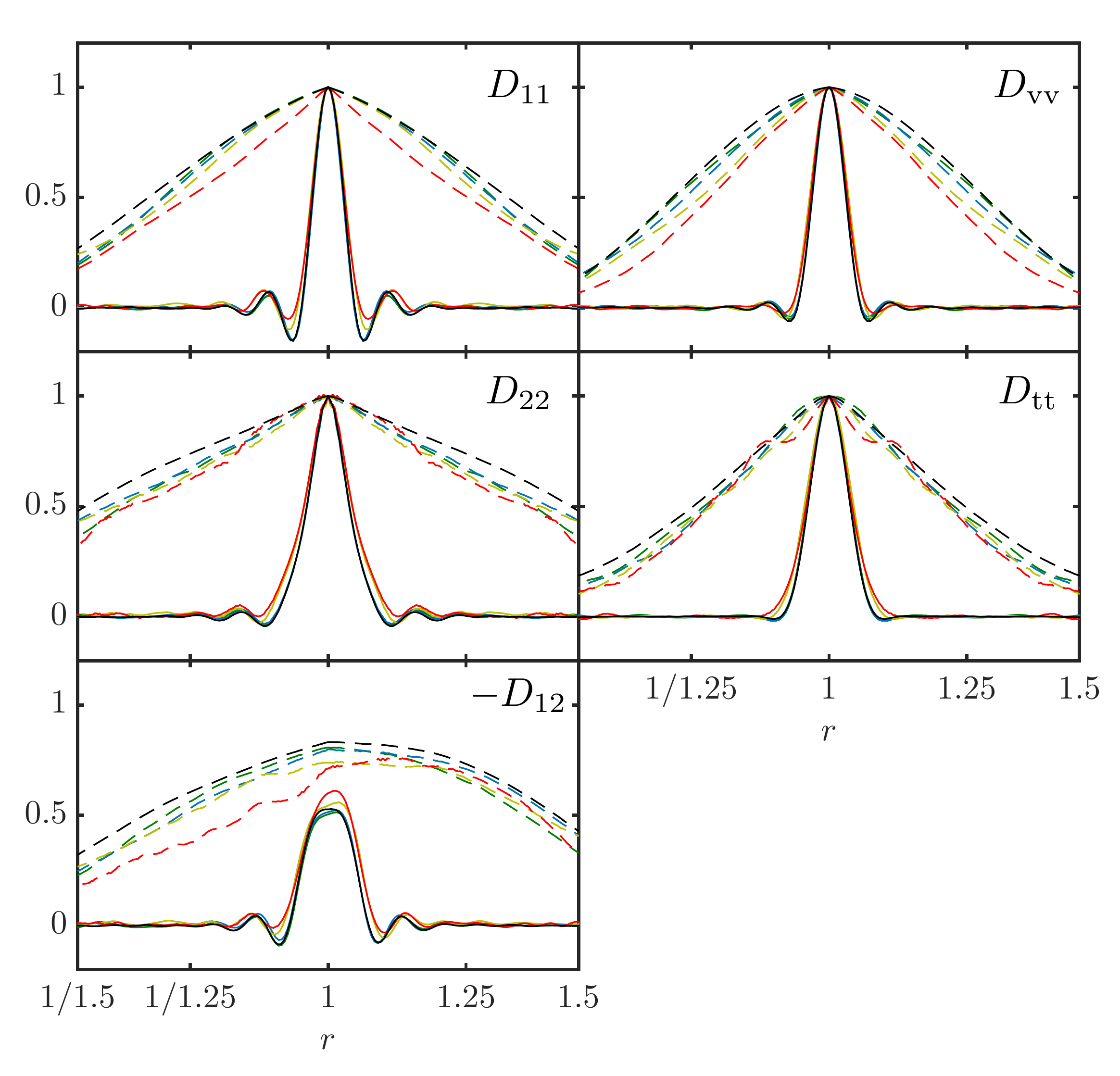}
\caption{Complete set of decoherence function in radiation era. Five different functions are shown: on the left $D_{11}$, $D_{22}$ and $D_{21}$  from top to bottom, and on the right $D_{\text{vv}}$ and $D_{\text{tt}}$ from top to bottom. The colour scheme is maintained from previous plots: red for $\beta=0.01$, yellow for $\beta=0.025$, blue for $\beta=0.1$ green for $\beta=0.25$ and black for $\beta=1.0$. The decoherence functions are computed for $k\tau=10.5$ (dashed lines) and $k\ta=100$ (solid lines). \label{Fig:CohFunrad}}
\end{figure}

In order to account for the different energy-density scale of each one of the $\beta$, we calculated the ETCs dividing them by the value $\mu^2(\beta)$, computed using the formulae (\ref{e:StrTen}) and (\ref{e:StrTenFun}). The results can be found in the panel ``b)" of Fig.~\ref{Fig:ETCrad} for radiation era (see Appendix for the corresponding matter era figure). The rescaling of the ETCs with $\mu^2$ brings the ETCs much closer to each other, almost lining them up (and for the tensor case, they do line up). However, there are still noticeable gaps  between different $\beta$, a clear evidence that the different normalizations do not depend only on the string tension. 

Nevertheless, it is interesting to note that  even though the ETCs from simulations with higher $\beta$ are still above those with lower $\beta$,  there is virtually no gap between the ETCs from the three lowest $\beta$. 
We interpret this as convergence of the $\mu^2$-rescaled ETCs at small $\be$.

We calculate the extra factor needed to match the ETC from lowest $\beta$ to that given by the case with $\beta=1$, and call that factor $\kappa^{\rm R,M}$:

\begin{equation}
E^{\rm R,M}_{a}(k\tau,\beta) = \kappa_a^{\text{R,M}} \left(\frac{\mu(\be)}{\mu(1)}\right)^2 E^{\rm R,M}_{a}(k\tau,1)\,.
\label{eq:ETCkappa}
\end{equation}

Table~\ref{table:kappa} contains the values of $\kappa$ for the whole set of simulations in radiation and in matter eras. In order to measure it, we fitted the region near the peak of the ETCs ($k\tau\in[9,60]$). We use the ETCs extracted at $\tSim=450$ and $\tSim=600$ for $\beta=1$ in radiation and matter respectively and $\tSim=660$ for $\beta<1$ simulations. The results from this procedure can be found in Table~\ref{table:kappa}.
 
\begin{table}[h!]
\resizebox{0.48\textwidth}{!}{
\renewcommand{\arraystretch}{1.2}
\begin{tabular}{|c|c|c|c|c||c|}
\hline
\multicolumn{6}{|c|}{$\kappa_a^{\text{R}}$}\\\hline
$\beta$ &$E_{11}$ & $E_{22}$ & $E_{12}$ & $E_{\rm vv}$ & Mean \\\hline
0.01 &0.65$\pm$0.06 & 0.53$\pm$0.05& 0.57$\pm$0.05 & 0.63$\pm$0.03 & 0.59$\pm$0.09 \\
 0.025 &0.58$\pm$0.03 & 0.51$\pm$0.03& 0.55$\pm$0.03 & 0.58$\pm$0.02 & 0.56$\pm$0.05 \\
 0.1 &0.64$\pm$0.02 & 0.59$\pm$0.03& 0.62$\pm$0.03 & 0.67$\pm$0.02 & 0.63$\pm$0.05 \\
 0.25 &0.86$\pm$0.03 & 0.84$\pm$0.03& 0.84$\pm$0.03 & 0.86$\pm$0.02 & 0.85$\pm$0.05 \\\hline
  \multicolumn{6}{|c|}{$\kappa_a^{\text{M}}$}\\\hline
  $\beta$ &$E_{11}$ & $E_{22}$ & $E_{12}$ & $E_{\rm vv}$ & Mean \\\hline
  0.01 &0.65$\pm$0.04 & 0.46$\pm$0.04& 0.49$\pm$0.05 & 0.56$\pm$0.03 & 0.54$\pm$0.08 \\
 0.025 &0.59$\pm$0.04 & 0.53$\pm$0.04& 0.55$\pm$0.03 & 0.58$\pm$0.03 & 0.56$\pm$0.07 \\
 \hline
\end{tabular}}
 \caption{\label{table:kappa} Values of the extra normalization factor $\kappa_a^{\text{R,M}}$ for different correlators (except tensors) and different cosmologies R(adiation) and M(atter). The values are obtained by fitting for the region around the peak or maximum of the correlator ($k\tau\in[9,60]$) and for ETCs extracted at $\tSim=450$ and $\tSim=600$ ($\beta=1$) in radiation and matter respectively and $\tSim=660$ for $\beta<1$ simulations. Errors on each correlator come from averaging over the contribution of all bins in $k\tau\in[9,60]$.}
\end{table}

\begin{figure}[h!]
\centering
\includegraphics[width=0.5\textwidth]{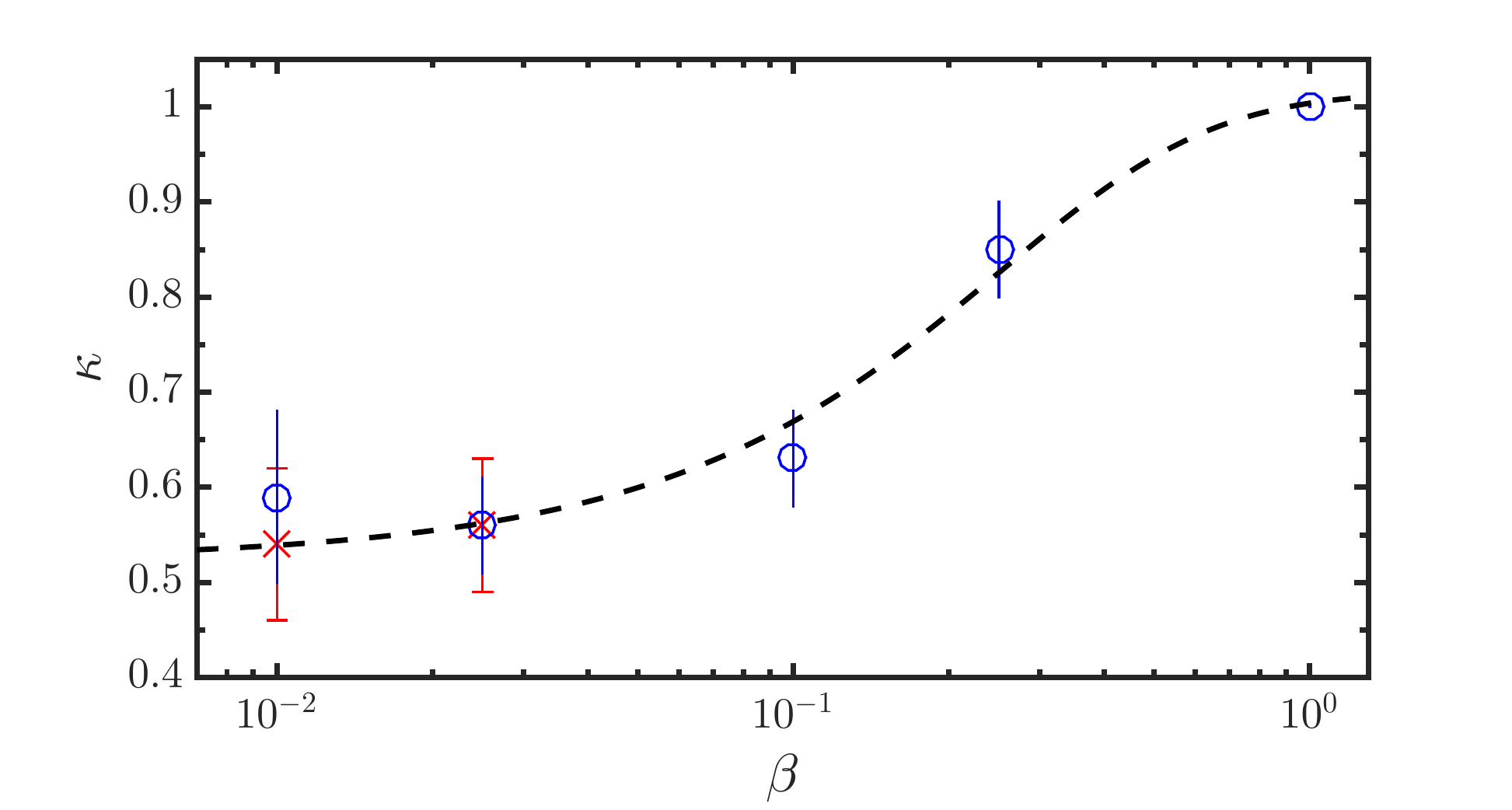}\\
\caption{Mean values and errors of the measure of $\kappa_a^{\text{R}}$ (blue) and $\kappa_a^{\text{M}}$ (red) as a function of $\beta$. The black dashed line corresponds to the function expressed in Eq. (\ref{eq:kappabeta}).
\label{Fig:kappa}}
\end{figure}
 
The mean points and the standard deviations for each $\beta$ are represented in Fig.~\ref{Fig:kappa}, where we use blue for the data from radiation simulations and red for matter simulations. This figure shows that for the lowest $\beta$'s the value of the factor seems to tend to an asymptote. We obtained an analytical expression for $\kappa(\beta)$ fitting those points and using the following function:

\begin{equation}
\kappa(\be) = \kappa_0 + a \tanh(b\log(\beta+1)) \quad (\be < 1),
\label{eq:kappabeta}
\end{equation}
where: $a=0.49\pm0.05$, $b=7.4\pm1.6$ and $\kappa_0=0.52\pm0.03$.

 This is the functional form for interpolation for different values of $\beta$.  
The asymptotic value for any arbitrarily low $\beta$ is, therefore, $\kappa_0=0.52 \pm 0.03$, which is within $1\sigma$ confidence-limit of our measurement for our lowest $\beta$ simulations, suggesting that the $\kappa(\be)$ correction for lower values of $\beta$ is of $O(1)$.
Tensor correlators do not require any extra normalization beyond $\mu^2$, 
and it is consistent to take $\ka^\text{R,M}_\text{tt} = 1$.


\section{$C_\ell$ and constraints} \label{sec:cls}

Obtaining the CMB anisotropies power spectra from strings is rather involved. The procedure and high resolution $C_\ell$'s from $4096^3$ simulations at $\beta=1$ can be found  in \cite{Lizarraga:2016onn}.

The fact that the ETCs from different $\beta$ are (almost) a rescaled version of each other prompts us to provide a recipe to give an approximate CMB anisotropies power spectra from strings at any $\beta$ $(<1)$, by using the aforementioned  high resolution $C_\ell$'s from $\beta=1$ \cite{Lizarraga:2016onn}, rather than having to compute the $C_\ell$'s for every different $\beta$.

The CMB power spectra from strings are linear responses to the unequal time correlators (UETCs) (\ref{emtens}). In the previous section we demonstrated that, to a good approximation, the UETCs at low $\beta$ can be 
obtained by rescaled those at $\beta=1$. Thus, our claim is that the $C_\ell$s for a given $\beta$ can be obtained by just rescaling the $C_\ell$s at $\be=1$ \cite{Lizarraga:2016onn}. Given the different scaling properties of the tensor source, this rescaling should be performed separately for the scalar, vector and tensor perturbation modes, or
\begin{equation}
C^{\text{L}}_{\ell}(\beta) = \kappa^{\text{L}}(\beta) \left(\frac{\mu(\be)}{\mu(1)}\right)^2 C^{\text{L}}_{\ell}(1)\,,
\label{eq:clscaling}
\end{equation}
where $L = S,V,T$ and 
\bea
\kappa^{\text{S,V}}(\beta)&=&\kappa(\beta) \\
\kappa^\text{T}(\beta) & = & 1.
\eea
We recall that the function $\kappa(\beta)$ is given in  Eq.~(\ref{eq:kappabeta}), and 
$\mu(\be)/\mu(1) = B(\beta)$,
which is approximated in (\ref{e:StrTenFun}).
We can then estimate the  CMB constraints  to any $\beta<1$. 

So far the constraints imposed by CMB experiments, such as Planck, in the abundance of cosmic strings come from channels (TT, TE and EE) where tensor modes are the least dominant. Therefore, it is reasonable to assume that the entire normalisation difference can be encapsulated in $\kappa(\beta)=\kappa^{\text{S,V}}(\beta)$, and that the mismatch with $\kappa^{\text{T}}$ produces a negligible effect in the new constraints.

 For example, we can obtain the constraints in $G\mu$ for Planck 2015 as a function of $\beta$, by just extrapolating our $\beta=1$ results \cite{Lizarraga:2016onn}, which imply that $\kappa^{1/2}(\beta)G\mu(\beta)  \lesssim 2.0\times 10^{-7}$.   Oour results suggest that $\kappa(\beta)$ would be of order $O(1)$ for lower values of $\beta$. 
This information can be recast into obtaining constraints for the energy scale of the symmetry breaking $\phi_0$ as a function of $\beta$: 
 
\begin{eqnarray}
  \phi_0  \kappa(\beta)^{1/4}&\lesssim& B(\beta)^{-1/2} 2.2 \times 10^{15}\,\text{GeV}\,,
\end{eqnarray}
with $\kappa(\beta)\sim O(1)$ for very low $\beta$. 
This shows that models with   low $\beta$ would allow higher symmetry-breaking scales to be  compatible with the constraints on strings from the CMB. 
 

\section{Conclusion and Discussion} \label{sec:conclusions}

In this paper we have presented the first network simulations of the Abelian Higgs model for $\beta < 1$
to include both matter and radiation eras and Cosmic Microwave Background (CMB) constraints.
The comoving mean string separation $\xi(\tau)$ is found to 
grow in a manner consistent with linear,  giving evidence  that the strings are scaling as usual. The dependence of the growth rate $d\xi / d\tau$ on $\beta$ is weak, and consistent with it being independent of $\beta$.  
The lack of dependence of $d\xi/d\tau$ on $\beta$ is more evidence that the radiative energy-loss mechanism studied in \cite{Hindmarsh:2017qff} is non-perturbative: perturbative scalar radiation should decrease with the scalar self-coupling $\lambda$ and hence $\beta$.

Our result for the mean string separation differs from that of \cite{Hiramatsu:2013tga}, who found that the mean string separation 
towards the end of the simulations 
reached a minimum at $\beta = 0.4$, and increased again at the lowest $\beta$ achieved, $\beta = 0.2$.  
Our initial conditions are prepared with a period of cooling to remove short-distance oscillations and bring the scalar field 
as close as possible to its minimum, consistent with there being a string network.  
The simulations of \cite{Hiramatsu:2013tga} on the other hand do not have cooling, and show signs of a coherent oscillation of the scalar field, 
which affects the string length measurement. 
Our simulations also reach string separations which are significantly larger. 
We therefore agree with the tentative conclusions of \cite{Hiramatsu:2013tga} that the apparent $\be$-dependence of the string separation should 
disappear for larger simulations.

The stress-energy correlators $C_{ab}$ on the other hand do depend on $\beta$.
This dependence is well described by a simple rescaling with the string mass with the string tension $\mu$, $C_{ab} \propto (\mu(\beta)/\mu(1))^2$, except for an additional O(1) correction that we call $\kappa$, as given in Eq.\ (\ref{eq:ETCkappa}). We measure this correction factor in our simulations and find that the values for all UETCs except for the tensors are comparable.  A functional fit to our measured $\kappa$ values for any $\beta$ is given in Eq.\ (\ref{eq:kappabeta}) showing that for $\beta \lesssim 0.1$ it asymptotes to $\kappa_0 = 0.52 \pm 0.03$, suggesting $\kappa~O(1)$ for very low values of $\beta$. For the tensor-type stress-energy UETCs we find no additional correction, i.e.\ $\kappa_{tt} \simeq 1$. As the tensor UETCs do not depend on $T^{0i}$, while the other UETCs do, we conjecture that the correction factor, which encodes the effect of the string interactions when $\beta\neq1$, may reflect that networks with lower values $\beta$ have slightly smaller velocities.

The simple scaling of the scalar and vector UETCs implies that also the angular power spectrum of temperature fluctuations $C^{TT}_\ell$ is simply rescaled by the same factor, given in Eq.\ (\ref{eq:clscaling}), as the tensor UETCs contribute to the temperature fluctuations only at a low level. This equation allows to predict the CMB angular anisotropy power spectra for any $\beta<1$. Because of this rescaling, models with low (effective) self-coupling can satisfy the  CMB bounds at higher symmetry-breaking scales.

If our results can be extrapolated to very low $\beta$, they would allow higher symmetry-breaking scales to be  compatible with the constraints on strings from the CMB.  
This is particularly relevant for models where the symmetry is broken along a flat direction lifted by supersymmetry-breaking terms \cite{Lazarides:1986di}, which can have effective values of $\beta$ in the range $10^{-30} \lesssim \beta \lesssim 10^{-15}$ \cite{Cui:2007js}.  The supersymmetric F-term hybrid inflation model of \cite{Hindmarsh:2012wh} has $\be_\text{eff} \sim 10^{-24}$, and so $B \simeq 0.04$. When normalised to the amplitude of inflationary perturbations, it is found\footnote{We use the asymptotic value of $\kappa$ to obtain concrete values of  $G\mu$, though given that the values of $\beta$ are extremely low these numbers should be interpreted very cautiously.} that $G\mu \simeq 10^{-7}$, compatible with current CMB constraints.  
Strings formed at $10^{16}$ GeV (around the supersymmetric GUT scale) would be allowed if $\beta \lesssim 10^{-15}$, although to reach $2\times 10^{16}$ GeV would need $\beta \lesssim 10^{-62}$.
Such low-$\beta$ models must also find ways to avoid the complementary $\ga$-ray \cite{Mota:2014uka} or gravitational wave \cite{Ringeval:2017eww,Blanco-Pillado:2017rnf} constraints.  One possibility is that the string network decays entirely into particles in a hidden sector.

  
\begin{acknowledgments}
 We thank the authors of Ref.~\cite{Hiramatsu:2013tga} for helpful comments and discussion.
This work has been supported by two grants from the Swiss National Supercomputing Centre (CSCS) under Project IDs s319 and s546. In addition, this work has been possible thanks to the computing infrastructure of the i2Basque academic network, the COSMOS Consortium supercomputer (within the DiRAC Facility jointly funded by STFC and the Large Facilities Capital Fund of BEIS), and the Andromeda/Baobab cluster of the University of Geneva. 
MH (ORCID ID 0000-0002-9307-437X) acknowledges support from the Science and Technology Facilities Council 
(Grant No. ST/P000819/1).
DD (ORCID ID 0000-0002-4693-4891) and 
MK (ORCID ID 0000-0002-3052-7394) acknowledge support from the Swiss NSF.
JL (ORCID ID 0000-0002-1198-3191) and 
JU (ORCID ID 0000-0002-4221-2859) acknowledge support from Eusko Jaurlaritza (IT-979-16) and the Spanish Ministry MINECO (FPA2015-64041-C2-1P).
 \end{acknowledgments}
 

\appendix

\section{Results from matter-era simulations}
\label{Sec:AppendixA}

\begin{figure*}
\centering
\includegraphics[width=0.49 \textwidth]{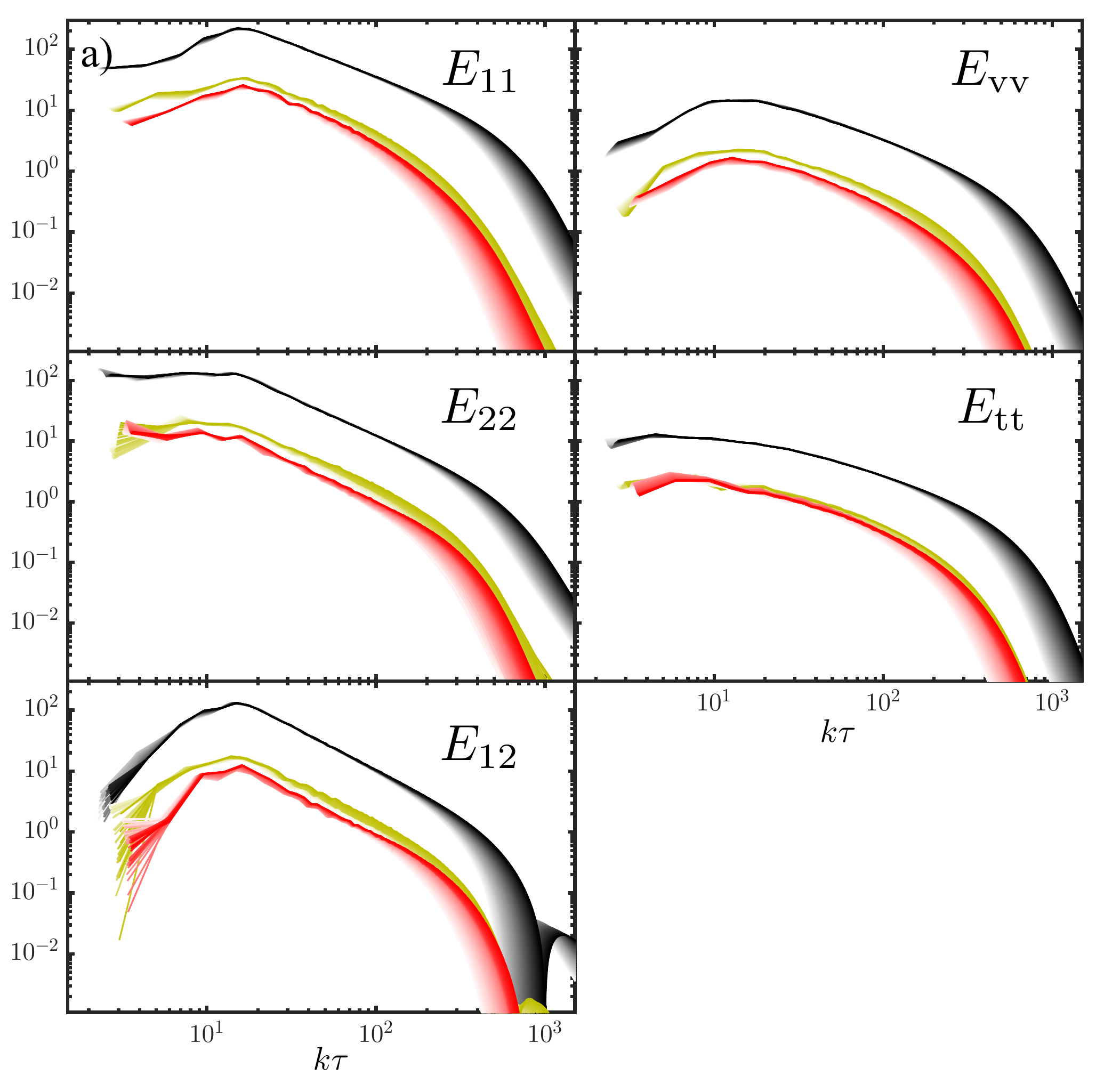}
\includegraphics[width=0.49\textwidth]{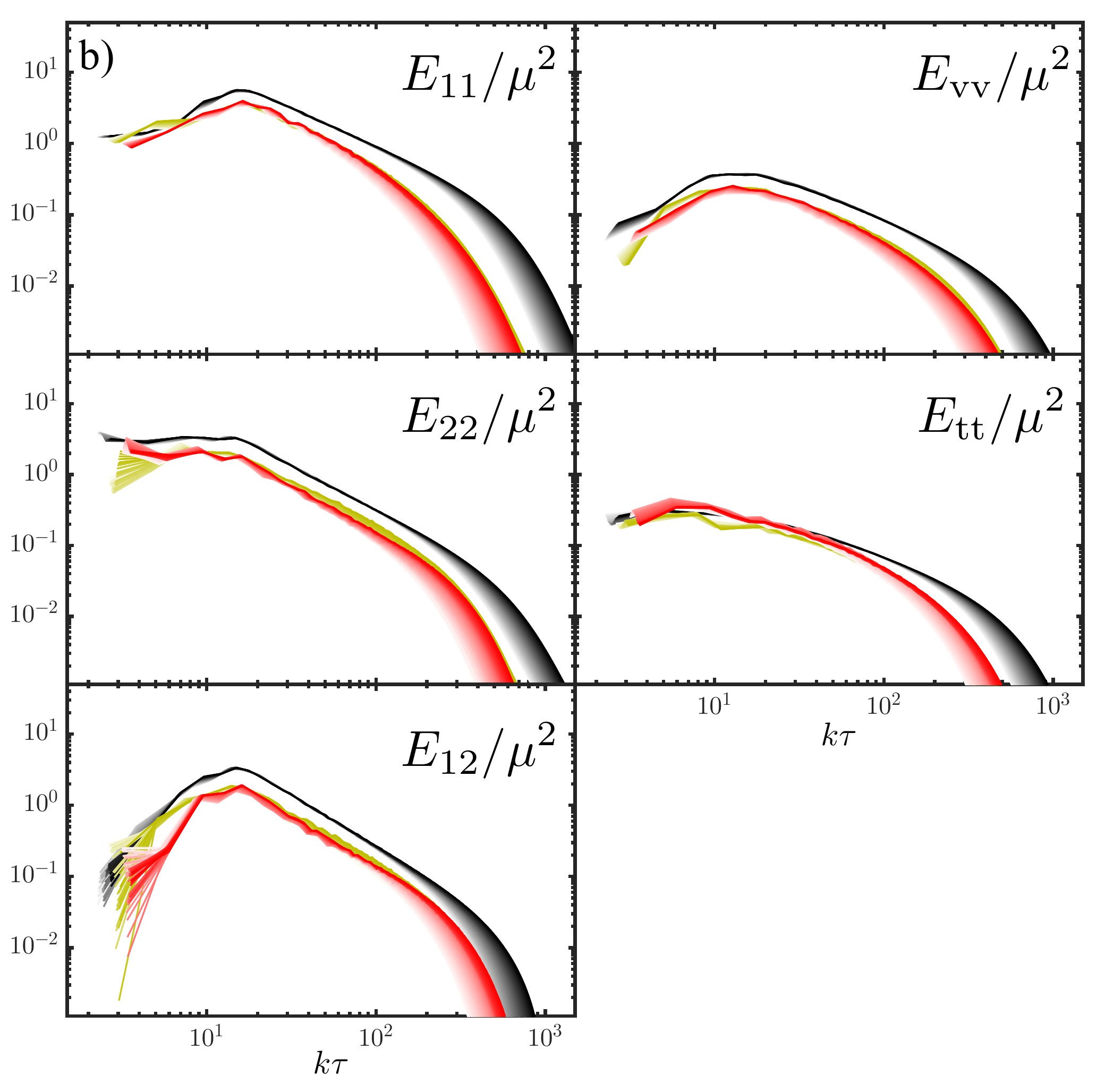} 
\caption{As Fig. \ref{Fig:ETCrad} but with the ETCs from matter domination. ETCs are extracted from $\tSim\in[600, 700]$ for $\beta<1$ and $\tSim\in[600, 750]$ for $\beta=1$.\label{Fig:ETCmat}}
\end{figure*}

\begin{figure}
\centering
\includegraphics[width=0.49\textwidth]{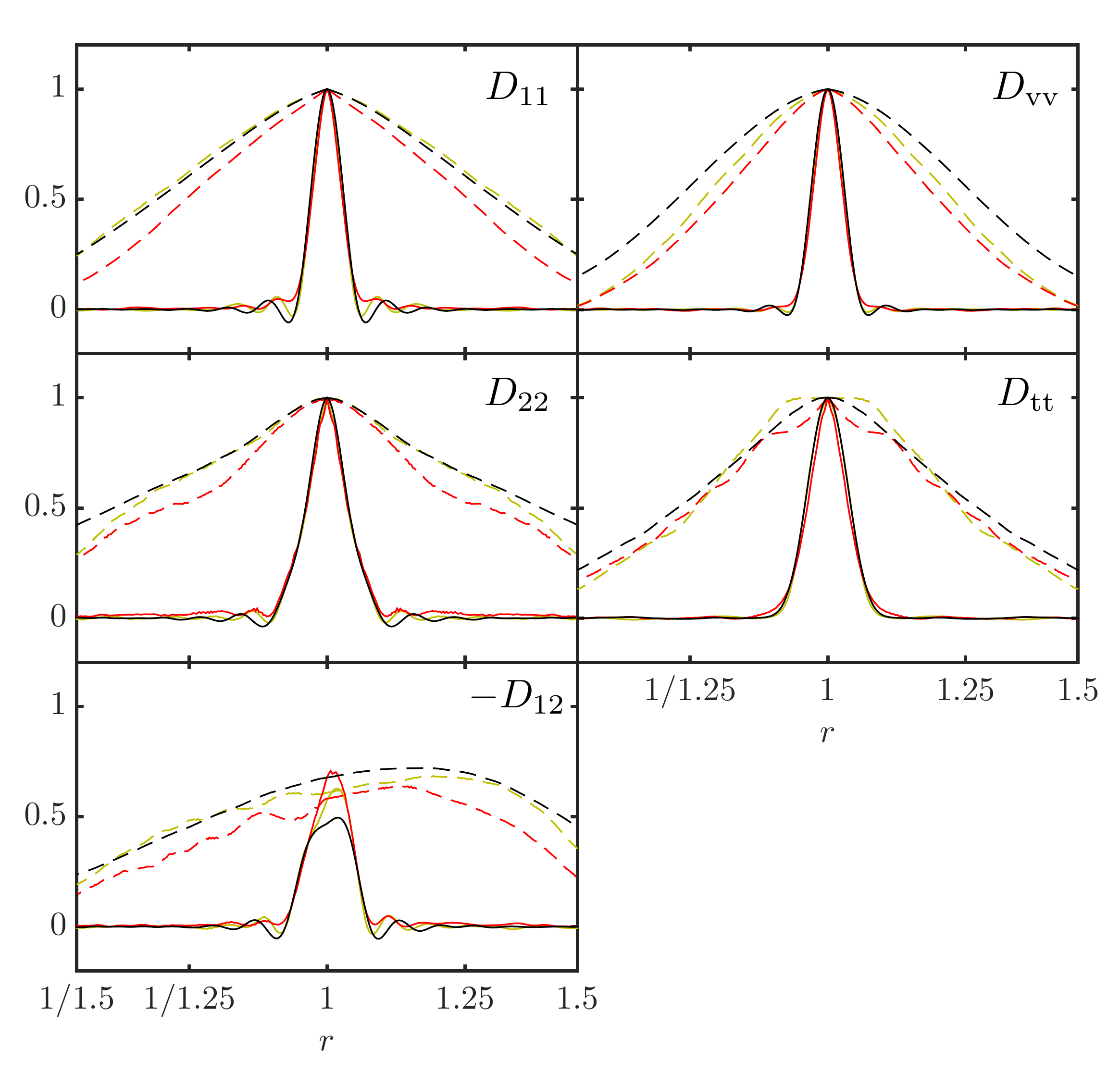} 
\caption{Complete set of decoherence function in matter era.  Five different functions are shown: on the left $D_{11}$, $D_{22}$ and $D_{21}$  from top to bottom, and on the right $D_{\text{vv}}$ and $D_{\text{tt}}$ from top to bottom. The colour scheme is maintained from previous plots: red for $\beta=0.01$, yellow for $\beta=0.025$, blue for $\beta=0.1$ green for $\beta=0.25$ and black for $\beta=1.0$. The decoherence functions are computed for $k\tau=10.5$ (dashed lines) and $k\ta=100$ (solid lines). \label{Fig:CohFunmat}}
\end{figure}

Here we show the equal time correlators (Fig.~\ref{Fig:ETCmat}) and decoherence functions (Fig.~\ref{Fig:CohFunmat}) from the three matter era simulations, which complement the radiation era data show in Figs.~\ref{Fig:ETCrad}, \ref{Fig:CohFunrad}. The parameters of the runs are given in Table \ref{table:param}.

\bibliography{CosmicStrings}

\end{document}